\documentclass[10pt,aps,superscriptaddress,twocolumn,amsmath,amssymb,floatfix,prb,nofootinbib]{revtex4-2}

\usepackage[T1]{fontenc}
\usepackage[utf8]{inputenc}
\usepackage[colorlinks, linkcolor=mycol, citecolor=mycol, urlcolor=mycol, breaklinks]{hyperref}
\usepackage{braket}
\usepackage{bbold}
\usepackage{enumerate}
\usepackage{tikz}
\usetikzlibrary{mindmap,arrows.meta, decorations.markings}
\usepackage{graphicx}
\usepackage{dcolumn}
\usepackage{bm}
\usepackage{physics}
\usepackage{xcolor}
\usepackage{units}
\usepackage{comment}
\usepackage{times}
\usepackage[normalem]{ulem}
\usepackage{cancel}

\definecolor{mycol}{RGB}{10,55,130}

\DeclareMathAlphabet\mathbfcal{OMS}{cmsy}{b}{n}

\begin{document}

\title{Microscopic quantum description of surface plasmon polaritons: \\ Revealing intrinsic ultrastrong light-matter coupling}

\author{Florian Maurer}
\affiliation{Universit\'{e} de Strasbourg, CNRS, Institut de Physique et Chimie des Matériaux de Strasbourg, UMR 7504, F-67000 Strasbourg, France}
\author{Thomas F. Allard}
\affiliation{Universit\'{e} de Strasbourg, CNRS, Institut de Physique et Chimie des Matériaux de Strasbourg, UMR 7504, F-67000 Strasbourg, France}
\affiliation{Departamento de F\'{i}sica Te\'{o}rica de la Materia Condensada, Universidad Aut\'{o}noma de Madrid, E-28049 Madrid, Spain}
\affiliation{Condensed Matter Physics Center (IFIMAC), Universidad Aut\'{o}noma de Madrid, E-28049 Madrid, Spain}
\author{Yanko Todorov}
\email{yanko.todorov@espci.fr}
\affiliation{Laboratoire de Physique et d'\'{E}tude des Mat\'{e}riaux, LPEM, UMR 8213, ESPCI Paris, Universit\'{e} PSL, CNRS, Sorbonne Universit\'{e}, F-75005 Paris, France}
\author{Guillaume Weick}
\email{guillaume.weick@ipcms.unistra.fr}
\affiliation{Universit\'{e} de Strasbourg, CNRS, Institut de Physique et Chimie des Matériaux de Strasbourg, UMR 7504, F-67000 Strasbourg, France}
\author{David Hagenm\"{u}ller}
\email{david.hagenmuller@ipcms.unistra.fr}
\affiliation{Universit\'{e} de Strasbourg, CNRS, Institut de Physique et Chimie des Matériaux de Strasbourg, UMR 7504, F-67000 Strasbourg, France}


\begin{abstract} 
We develop a microscopic quantum theory of surface plasmon polaritons valid for arbitrary metal-dielectric geometries. Our framework is based on the Power-Zienau-Woolley representation of quantum electrodynamics, which provides an optimal separation between electronic and photonic degrees of freedom and is therefore particularly well suited for constructing quantum descriptions of polaritonic excitations in strongly dispersive media. Within this formulation, the fundamental electronic oscillator is identified as the bulk plasmon mode, which is nonperturbatively coupled to the radiative continuum of free photon modes. This coupling induces a geometry-dependent renormalization of the bulk plasma frequency, giving rise to confined plasmonic resonances. As specific applications, we recover the localized surface plasmon modes of metallic nanoparticles, including radiative frequency shifts and decay, as well as the exact dispersion relation of propagating surface plasmon polaritons at planar interfaces. Our quantum treatment further reveals that light-matter interactions at metal-dielectric interfaces are inherently in the ultrastrong coupling regime. As a result, in the quasistatic limit, the system exhibits unconventional ground-state quantum fluctuations that can be controlled through the refractive index. These results open new intriguing perspectives in the field of quantum plasmonics.
\end{abstract}

\maketitle

\section{Introduction}

Surface plasmon polaritons are hybrid excitations arising from the strong coupling between the electromagnetic (EM) field and collective electron oscillations at metal-dielectric interfaces. They are central to nanophotonics, enabling the confinement of light at subwavelength scales, far beyond the diffraction limit~\cite{Raether1988,Barnes2003,Biagioni_2012}. Two primary types are commonly distinguished: propagating surface plasmon (PSP) polaritons, which travel along interfaces, and localized surface plasmons (LSPs), confined to metallic nanoparticles. Such confinement facilitates access to extreme light-matter interaction regimes by coupling surface plasmon polaritons with excitons or phonons~\cite{Torma_2015}, forming hybrid eigenstates, often termed plexcitons. The strong coupling regime, where the vacuum Rabi splitting between plexciton modes exceeds dissipation rates, has been demonstrated in systems including J-aggregates~\cite{Bellessa2004,Dintinger2005,Sugawara2006,Guebrou2012}, dye molecules~\cite{Pockrand1982,Hakala2009}, quantum dots~\cite{Gomez2010,Hoang2016,Santhosh2016}, and two-dimensional materials~\cite{Koppens2011GraphenePlasmonics,Liu2016,Zheng2017,Kleemann2017,Stuhrenberg2018,Andolina2025QEDGrapheneLandau}.

More recently, attention has shifted toward nonperturbative regimes like ultrastrong and deep strong coupling, where the interaction strength becomes comparable to the transition energy~\cite{Ciuti2005,Forn-Diaz2019,kockum2019ultrastrong,Qin2024}. In the field of plasmonics, such regimes have been achieved with molecular aggregates on plasmonic structures~\cite{Balci2013,Todisco2018} and in nanoparticle arrays~\cite{ baranov2020ultrastrong, lamow18_PRB, mueller2020deep}. Strong light-matter interactions have enabled macroscopic quantum phenomena such as polariton condensation~\cite{kasprzak2006bose} and superfluidity~\cite{amo2009superfluidity,Carusotto2013QuantumFluids}. Over the past decades, this has led to the emergence of polaritonics~\cite{garcia-vidal2021manipulating,basov2025polaritonic}, a field exploring how coupling to vacuum fields can modify material properties, including, for example, quantum transport~\cite{paravicini2019magneto,appugliese2022breakdown,faist2025tunable}.

In this context, a microscopic quantum model capable of describing quantum emitter-surface plasmon polariton interactions at arbitrarily large coupling strengths is highly desirable.
Owing to their metallic constituents, plasmonic structures inherently exhibit strong dispersion and absorption, as imposed by Kramers-Kronig relations.
Quantizing the EM field in dispersive and dissipative media has long posed a challenge in quantum electrodynamics (QED). A major advance was the extension of the Hopfield polariton model~\cite{Hopfield} to reservoir-based approaches~\cite{Huttner1992,Drummond1999,Suttorp2004,Gubbin2016}, where dissipation is treated via a bath of harmonic oscillators coupled to the material polarization, itself interacting with the free EM field. Formal diagonalization via the Fano method~\cite{Fano1956} yields polaritonic modes that act as noise sources for the EM field through the Green tensor~\cite{Imam1994,Dung1998,Waks2010,Semin2025}, which contains the full field information but is often too complex for direct computation. Expansions in terms of a few quasinormal modes offer a computationally efficient alternative in simple geometries~\cite{Garraway1997,Franke2019}. These ideas are unified in the macroscopic QED framework~\cite{scheel2008macroscopic,Feist2021}, formulated in the Power-Zienau-Woolley (PZW) representation, which provides a rigorous, though numerically sensitive, foundation for quantum optics in complex media.

Effective models that quantize the PSP polariton field as a single degree of freedom have been proposed~\cite{Tame2008,Archambault2010,Tudela2013}. However, PSP polaritons are intrinsically hybrid light-matter excitations, implying that the associated field is not purely transverse but also contains a longitudinal component. As a consequence, such effective descriptions cease to be valid in the ultrastrong-coupling regime when quantized PSP polaritons interact with additional quantum degrees of freedom~\cite{Hagenmuller2019}. Other approaches that neglect losses have proposed more consistent quantization schemes for PSP polaritons by treating them as hybrid modes~\cite{Alpeggiani,Todorov2014}. Nevertheless, the formalism used in Ref.~\cite{Alpeggiani}, based on the minimal coupling representation, requires \textit{a priori} knowledge of the solution and becomes cumbersome due to the $\textbf{A}^2$ term~\cite{Allard}, which acquires a complicated form as a result of the broken translational symmetry perpendicular to the metal-dielectric interface. Meanwhile, the approach of Ref.~\cite{Todorov2014} proposes a nonstandard framework in which bulk metal plasma oscillations are coupled self-consistently to a heuristic, geometry-dependent evanescent photon field.

In this work, we introduce a microscopic quantum description of surface plasmon polaritons valid for arbitrary geometries. To retain analytical tractability, we neglect nonradiative losses while preserving the essential physics. This guarantees that the model can be consistently extended to describe coupling to additional quantum degrees of freedom beyond the perturbative regime of quantum electrodynamics, without introducing unphysical artifacts.

Our formulation starts from the full Lagrangian in the PZW representation~\cite{Power1959,Woolley1971}, whose advantages will be discussed below. The PZW description relies on the knowledge of the polarization field of free electrons, which is provided here onto a discrete mode basis derived from the solutions of Laplace's equation. This polarization field is coupled to the displacement field of the free EM field. The quantization procedure is rigorously defined, allowing us to explicitly determine the plasmon eigenmodes of the full Hamiltonian, and to define their corresponding quantization lengths. In our approach, the fundamental electronic excitation is identified as the bulk plasmon of the metal at frequency $\omega_{\mathrm{p}}$, which is coupled to a continuum of free-space radiation modes. The latter is nonperturbatively modified by the geometry of the metallic region. Within this framework, we recover the LSP modes of metallic nanoparticles, including radiative decay and frequency renormalization, as well as the surface plasmon mode frequency of planar metal-dielectric interfaces. Starting from these quasistatic excitations, PSP polaritons in layered geometries could be obtained. Here, we illustrate this approach for a planar interface, for which we recover the exact dispersion relation of PSP polaritons.

Being intrinsically quantum, our approach provides direct access to the photonic and electronic contributions to the polaritonic modes, as well as to the plasmonic population of the polaritonic ground state in the quasistatic regime. This population is finite, demonstrating that plasmon-photon coupling at metal-dielectric interfaces is inherently in the ultrastrong coupling regime. Moreover, it depends solely on the interface geometry and the refractive indices of the media. For typical materials, the ground-state plasmonic population can reach values of order unity.

Within this framework, the radiative decay of LSPs is naturally recovered in a fully quantum-mechanical manner. To this end, we employ the formalism of electronic Green's functions, widely used in the description of microscopic devices~\cite{Datta_1995}. Although in this case our approach would likely reduce to macroscopic QED, we emphasize that nonradiative losses can be incorporated either within the Green's function formalism or through Fano diagonalization~\cite{Fano1961}, by coupling the electronic degrees of freedom to additional baths of quantum harmonic oscillators.

The paper is organized as follows: In Sec.~\ref{sec_electro}, we develop the general Lagrangian and Hamiltonian formalism to describe arbitrary metal-dielectric interfaces within the PZW framework, and introduce a quantization procedure for the polarization field in the metal. In Sec.~\ref{Sec_cases}, we apply this formalism to two representative cases: LSP polaritons in a spherical nanoparticle (Sec.~\ref{sec : LSP qs}) and PSP polaritons at a planar interface (Sec.~\ref{sec : SP qs}). Conclusions and perspectives are presented in Sec.~\ref{conclusion}.

\section{Electrodynamics of metal-dielectric interfaces in the PZW framework}
\label{sec_electro}

\begin{figure}[t!]
\includegraphics[scale=0.65]{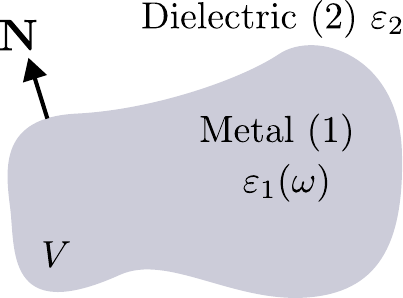}
\caption {Schematic representation of a metal-dielectric interface. The system comprises a metallic region (1) of volume $V$, assumed to exhibit negligible nonradiative losses and described by a Drude-type dielectric function $\varepsilon_{1}(\omega)=\varepsilon_{\infty}- \omega^{2}_{\mathrm{p}}/\omega^2$, where $\varepsilon_{\infty}$ denotes the background dielectric constant, and $\omega_{\mathrm{p}}$ is the plasma frequency. Adjacent to this is a dielectric region (2), characterized by a positive dielectric constant $\varepsilon_2 >0$. The unit vector $\mathbf{N}$ is oriented perpendicular to the interface, pointing into the dielectric medium.} 
\label{fig : Interface}
\end{figure}

We consider a general system comprising a metallic region of volume $V$ embedded within a dielectric medium, as illustrated in Fig.~\ref{fig : Interface}. The metallic domain hosts a homogeneous electron gas of density $\rho$ immersed in a uniform, positively charged background, a configuration commonly referred to as jellium, ensuring overall charge neutrality. The electrons are characterized by a mass $m$ and a charge $-e$. Throughout this work, we limit our analysis to linear and nonmagnetic materials.

A central tool employed in our analysis is the Helmholtz decomposition of vector fields, which we write as $\mathbf{F}(\mathbf{r})=\mathbf{F}_{\parallel}(\mathbf{r})+\mathbf{F}_{\perp}(\mathbf{r})$. Here, $\mathbf{F}_{\perp}(\mathbf{r})$ denotes the transverse (solenoidal) component, satisfying $\bm{\nabla} \cdot \mathbf{F}_{\perp} (\mathbf{r})= 0$, while $\mathbf{F}_{\parallel}(\mathbf{r})$ is the longitudinal (irrotational) component, obeying $\bm{\nabla} \times \mathbf{F}_{\parallel} (\mathbf{r})= \mathbf{0}$. A simple Fourier space analysis reveals that for any well-behaved vector fields $\mathbf{F}$ and $\mathbf{G}$, their transverse and longitudinal components satisfy the orthogonality relation
\begin{equation}
\int_{\mathbb{R}^3} \! \mathrm{d}^3\mathbf{r} \, \mathbf{F}_{\perp}(\mathbf{r}) \cdot \mathbf{G}_{\parallel} (\mathbf{r}) =0 .
\label{orth_relation}
\end{equation}
It is important to note that this orthogonality condition is, in general, nonlocal in real space. 

\subsection{Generic formulation}
\label{Sec_lagra_Hamil}

In the PZW representation, the Lagrangian and Hamiltonian are expressed in terms of the polarization field ${\mathbf{P}}$ and the electric displacement field ${\mathbf{D}}$. The total PZW Lagrangian is decomposed as $\mathcal{L}=\mathcal{L}_{\mathrm{el}} +\mathcal{L}_{\mathrm{ph}}+\mathcal{L}_{\mathrm{int}}$~\cite{Cohen,Babiker}, where
\begin{align}
\mathcal{L}_{\mathrm{el}} &= \frac{m\rho}{2} \int_{\mathbb{R}^3} \! \mathrm{d}^3\mathbf{r} \,  \mathbf{v}^2 (\mathbf{r})\xi(\mathbf{r})
 - \frac{\varepsilon_0}{2} \int_{\mathbb{R}^3} \! \mathrm{d}^3 \mathbf{r} \, \mathbf{E}_{\parallel}^2 \nonumber \\
&= \frac{1}{2 \varepsilon_0 \omega_{\mathrm{p}}^2} \int_{V} \mathrm{d}^3\mathbf{r} \, \dot{\mathbf{P}}^2
 - \frac{\varepsilon_0}{2} \int_{\mathbb{R}^3} \! \mathrm{d}^3 \mathbf{r} \, \mathbf{E}_{\parallel}^2
\label{eq:Lagrange_matter}
\end{align}
describes the electronic degrees of freedom. The first term represents the kinetic energy of the electron gas within the metallic volume, while the second corresponds to the instantaneous Coulomb interaction, dependent solely on the longitudinal component of the electric field. The latter, in turn, depends only on the relative positions of the charges \cite{Cohen}. Here, $\varepsilon_{0}$ is the vacuum permittivity, and $\xi(\mathbf{r})$ is a function equal to $1$ in the metallic region ($\mathbf{r} \in V$) and $0$ elsewhere. The plasma frequency is defined by $\omega_{\mathrm{p}} = \sqrt{\rho e^2/m\varepsilon_0}$. The transition from the first to the second line in Eq.~\eqref{eq:Lagrange_matter} follows from the introduction of the electronic velocity field $\mathbf{v}(\mathbf{r})$ and the identification of the current density $\mathbf{j}(\mathbf{r}) = -e \rho \mathbf{v}(\mathbf{r})$, which relates to the polarization field via $\mathbf{j}(\mathbf{r})=\dot{\mathbf{P}}(\mathbf{r})$, with
$\dot{\mathbf{P}}\equiv \frac{\partial \mathbf{P}}{\partial t}$
\cite{Griffiths}. Outside the metallic region ($\mathbf{r} \notin V$), the polarization field vanishes, $\mathbf{P}(\mathbf{r}) = \mathbf{0}$, due to the absence of charge current in the dielectric.

The kinetic energy term in Eq.~\eqref{eq:Lagrange_matter} can be identified with the kinetic inductance of the metallic region, whereas the potential energy term corresponds to a capacitance associated with the distribution of quasistatic electric energy in the system. Accordingly, the electronic Lagrangian is formally equivalent to that of an inductor-capacitor circuit in which the inductance is purely of kinetic origin. As a result, our theory establishes a direct conceptual bridge between plasmonics and the fields of EM meta-atom resonators~\cite{Pendry_1999} and mesoscopic quantum circuits~\cite{Devoret1997}.

The Lagrangian of the free EM field is given by
\begin{equation}
\mathcal{L}_{\mathrm{ph}}=\frac{\varepsilon_0}{2}\int_{\mathbb{R}^3} \! \mathrm{d}^3 \mathbf{r} \, \left( \mathbf{E}^2_{\perp}-c^2 \mathbf{B}^2 \right), 
\label{eq:Lagrange_ph}
\end{equation}
where $c$ is the speed of light in vacuum, and $\mathbf{B} = \bm{\nabla} \times \mathbf{A}$ denotes the magnetic field, with $\mathbf{A}$ being the vector potential, constrained to be purely transverse under the Coulomb gauge condition $\bm{\nabla} \cdot \mathbf{A} = 0$. The interaction Lagrangian, dependent on the transverse electric field $\mathbf{E}_{\perp}$ and the polarization, reads
\begin{equation}
\mathcal{L}_{\mathrm{int}}=\int_{V} \! \mathrm{d}^3 \mathbf{r} \,\mathbf{P} \cdot \mathbf{E}_{\perp}
\label{eq:Lagrange_int}
\end{equation}
and includes retardation effects.

Given that the total system is charge-neutral, there are no external charges, and the displacement field satisfies $\bm{\nabla} \cdot \mathbf{D} = 0$, implying that $\mathbf{D}$ is purely transverse: $\mathbf{D} = \mathbf{D}_{\perp}$. The electric field decomposes as $\mathbf{E} = -\bm{\nabla} \phi - \dot{\mathbf{A}}$, with $\phi$ the electrostatic potential. Thus, one can identify $\mathbf{E}_{\parallel} = - \bm{\nabla} \phi$ and $\mathbf{E}_{\perp} = - \dot{\mathbf{A}}$. Substituting this into Eqs.~\eqref{eq:Lagrange_ph} and \eqref{eq:Lagrange_int}, the corresponding PZW Hamiltonian is obtained via a Legendre transformation with respect to $\dot{\mathbf{A}}$ and $\mathbf{v}$ \cite{Cohen}:
\begin{equation}
\label{legendre_H}
H=\int_{\mathbb{R}^3} \! \mathrm{d}^3 \mathbf{r} \left( \dot{\mathbf{A}} \cdot \frac{\delta \mathcal{L}}{\delta\dot{\mathbf{A}}} + \mathbf{v} \cdot \frac{\delta \mathcal{L}}{\delta\mathbf{v}}\right) - \mathcal{L}.
\end{equation}
The displacement field is defined as $\mathbf{D}=\varepsilon_0 \mathbf{E} +\mathbf{P}$. Using that $\mathbf{D}$ is purely transverse, the longitudinal components of the polarization and electric fields are related via $\mathbf{P}_{\parallel}= - \varepsilon_0 \mathbf{E}_{\parallel}$. Consequently, the total Hamiltonian \eqref{legendre_H} takes the form $H=H_{\mathrm{el}}+H_{\mathrm{ph}}+H_{\mathrm{int}}$, where the matter Hamiltonian is
\begin{equation}
\label{eq : Hmat 1}
    H_{\mathrm{el}} = \frac{1}{2 \varepsilon_0 \omega_{\mathrm{p}}^2}\int_{V} \mathrm{d}^3\mathbf{r} ~\dot{\mathbf{P}}^2   +  \frac{1}{2 \varepsilon_0} \int_{V} \mathrm{d}^3 \mathbf{r} \,  \mathbf{P}^2.
\end{equation}
This expression corresponds to the Hamiltonian of a collection of harmonic oscillators with a characteristic frequency equal to the \textit{bulk} plasma frequency $\omega_{\mathrm{p}}$.
The free EM Hamiltonian is given by
\begin{equation}
\label{eq : Hfield}
    H_\mathrm{ph} = \frac{1}{2 \varepsilon_0} \int_{\mathbb{R}^3} \mathrm{d}^3\mathbf{r} \, \left(\mathbf{D}^2 + c^2 \varepsilon_0^2\mathbf{B}^2\right),
\end{equation}
while the light-matter interaction term takes the form
\begin{equation}
\label{eq : Hint}
     H_\mathrm{int} = -\frac{1}{\varepsilon_0} \int_{V} \mathrm{d}^3\mathbf{r} \, \mathbf{P} \cdot \mathbf{D}.
\end{equation}

We have thus demonstrated that both the Lagrangian [Eqs.~\eqref{eq:Lagrange_matter}--\eqref{eq:Lagrange_int}] and the Hamiltonian [Eqs.~\eqref{eq : Hmat 1}--\eqref{eq : Hint}] can be expressed in terms of the electronic polarization field and the displacement field within the PZW representation. While the total polarization field, including both longitudinal and transverse components, appears in both terms of the matter Hamiltonian, only the longitudinal component contributes to the second term of the matter Lagrangian, as evident from Eq.~\eqref{eq:Lagrange_matter}.

Consequently, as will be shown in the following section, the characteristic frequency that emerges in the matter Lagrangian is not the bulk plasma frequency [unlike in the case of the matter Hamiltonian \eqref{eq : Hmat 1}] but instead corresponds to a surface plasmon frequency. This frequency arises from collective oscillations of the electron plasma localized near the interface and is renormalized relative to the bulk plasma frequency, with its value determined by the geometry of the metal-dielectric interface. To illustrate this, we apply our formalism to two specific configurations: a metallic spherical nanoparticle (Sec.~\ref{sec : LSP qs}) and a planar metal-dielectric interface (Sec.~\ref{sec : SP qs}).

Let us now discuss the advantages of the PZW representation over the conventional minimal-coupling formulation for condensed-matter problems such as the one addressed in the present work.
The key advantage of the PZW Hamiltonian lies in its structure: It contains a single light-matter interaction term, $H_{\mathrm{int}}$, while the remaining contributions ($H_{\mathrm{el}}$ and $H_{\mathrm{ph}}$) are strictly associated with the matter and photonic sectors, respectively. This stands in stark contrast with the minimal coupling Hamiltonian, wherein, alongside the bilinear interaction term $\propto \mathbf{p} \cdot \mathbf{A}$ (with $\mathbf{p}$ denoting the electron momentum) the quadratic term $\propto \int_{V} \mathrm{d}^3\mathbf{r} \, \mathbf{A}^2 (\mathbf{r})$ also contributes to the light-matter coupling, owing to the fact that the integration is confined to the spatial region occupied by the electrons. This additional coupling term introduces significant complications in the quantum treatment of metal-dielectric interfaces~\cite{Alpeggiani,Allard}. In contrast, as we will demonstrate, the PZW framework renders this treatment considerably more transparent and tractable. Note that the PZW formulation has been successfully applied to describe collective electronic excitations in confined electron gases and their interaction with light in the ultrastrong light-matter coupling regime~\cite{Todorov2010,delteil_charge-induced_2012}. These excitations are essentially plasmonic in nature.

The PZW approach based on the Hamiltonian $H = H_{\mathrm{el}} + H_{\mathrm{ph}} + H_{\mathrm{int}}$ has also previously been employed to construct a quantum theory of a homogeneous electron gas in unbounded space~\cite{Todorov2014}. In particular, it was shown that this framework yields the dielectric response of the metallic region in the form of a lossless Drude model,
\begin{equation}
\label{eq : Drude}
\varepsilon_{1}(\omega) = \varepsilon_{\infty} - \frac{\omega_{\mathrm{p}}^{2}}{\omega^{2}} .
\end{equation}
This dielectric function will be used in the following to define the boundary conditions at the metal surface. In the sequel, we neglect both nonradiative losses and nonlocal effects, i.e., we assume a purely real permittivity that is independent of the wavevector.

\subsection{Electronic degrees of freedom in the quasistatic regime}
\label{sec_quasilagrang}

By making use of the proportionality between the longitudinal components of the polarization and electric fields, it becomes evident that the electronic Lagrangian introduced in Eq.~\eqref{eq:Lagrange_matter} depends on the total polarization through the kinetic term, and specifically on its longitudinal component via the Coulomb interaction, i.e.,
\begin{equation}
\mathcal{L}_{\mathrm{el}} =\frac{1}{2 \varepsilon_0 \omega_{\mathrm{p}}^2} \int_{V} \mathrm{d}^3\mathbf{r} \, \dot{\mathbf{P}}^2 - \frac{1}{2\varepsilon_0} \int_{\mathbb{R}^3} \! \mathrm{d}^3 \mathbf{r} \, \mathbf{P}_{\parallel}^2.
\label{eq:Lagrange_matter2}
\end{equation}
Accordingly, a proper analysis of its properties requires the explicit identification of the two components $\mathbf{P}_{\parallel}$ and $\mathbf{P}_{\perp}$. As discussed in the preceding section, the total polarization is locally related to the current density and therefore vanishes outside the volume $V$. In contrast, the longitudinal component $\mathbf{P}_{\parallel}$ accounts for the effects of long-range Coulomb interactions and consequently extends significantly beyond the metallic region. Thus, the difference between the polarization field and its longitudinal component provides the physical richness of the plasmon modes confined by metallic interfaces.

We consider a globally neutral system, with no volume charges present in the dielectric medium, and assume that in the metallic region, charges reside only at the boundary surface of the metallic domain $\Sigma$. This is true if the typical size of the metallic region is much larger than the Thomas-Fermi screening length \cite{Kittel1963}. According to the Maxwell-Gauss equation, this implies that $\bm{\nabla} \cdot \mathbf{E} = \bm{\nabla} \cdot \mathbf{E}_{\parallel} = 0$ for all $\mathbf{r} \notin \Sigma$. Furthermore, using that $\mathbf{P}_{\parallel} = - \varepsilon_0\mathbf{E}_{\parallel}$ leads to $\bm{\nabla} \cdot \mathbf{P}_{\parallel} = 0$ for all $\mathbf{r} \notin \Sigma$. By definition, the longitudinal component also satisfies $\bm{\nabla} \times \mathbf{P}_{\parallel} = \mathbf{0}$. Therefore, the longitudinal polarization can be expressed as the gradient field 
\begin{equation}
\mathbf{P}_{\parallel} (\mathbf{r})= \varepsilon_0\bm{\nabla} \phi(\mathbf{r}).
\label{long_pola_electro}
\end{equation}
The electrostatic potential thus satisfies the Laplace equation, $\bm{\nabla}^2 \phi (\mathbf{r})= 0$, for all $\mathbf{r} \notin \Sigma$.

To solve Laplace's equation, appropriate boundary conditions must be specified, which depend on the geometry of the system and uniquely determine the potential $\phi$. First, $\phi$ must be continuous across the interface between the metal [region (1)] and the dielectric [region (2)]. Second, it must satisfy the boundary condition
\begin{equation}
\label{eq : BC Phi}
\varepsilon_1 (\omega) \frac{\partial \phi}{\partial \mathbf{N}}\Big|_1 = \varepsilon_2 \frac{\partial \phi}{\partial \mathbf{N}}\Big|_2,
\end{equation}
where $\varepsilon_1(\omega)$ denotes the relative permittivity of the metal from Eq.~\eqref{eq : Drude} , and $\varepsilon_2$ that of the dielectric medium. The unit vector $\mathbf{N}$ is normal to the interface and oriented outward into the dielectric region (see Fig.~\ref{fig : Interface}).

The general solution to Laplace's equation for the electrostatic potential can be expressed as
\begin{equation}
\label{eq : Phi Laplace}
\phi(\mathbf{r},t) = \sum_{\mu} \phi_{\mu}(t) g_{\mu}(\mathbf{r}),
\end{equation}
where $\mu$ is called separation index and reflects the symmetry of the system. The set of functions $g_{\mu}$ forms a complete orthonormal basis of harmonic functions, satisfying the orthogonality condition 
\begin{equation}
\int_{V} \mathrm{d}^3\mathbf{r}  \, g_{\mu}(\mathbf{r})  g^{*}_{\mu'}(\mathbf{r}) = \delta_{\mu, \mu'}
\end{equation}
within the metallic region. For the examples considered in this work, the gradient fields also obey the orthogonality condition
\begin{equation}
L^{2}_{\mu} \int_{V} \mathrm{d}^3\mathbf{r} \, \bm{\nabla} g_{\mu}(\mathbf{r}) \cdot \bm{\nabla} g^{*}_{\mu'}(\mathbf{r}) = \delta_{\mu, \mu'},
\label{effective_length}
\end{equation}
where $L_{\mu}$ is an effective mode length. This condition is, \textit{a priori}, not generally satisfied by the gradients of harmonic functions. However, 
one can always perform a unitary transformation to ensure that the condition in Eq.~\eqref{effective_length} is fulfilled.\footnote{It is not evident that the set of harmonic functions $g_{\mu}(\mathbf{r})$ satisfies the orthogonality condition \eqref{effective_length} for an arbitrary geometry. However, the matrix
$M_{\mu\mu'} = \int_{V} \mathrm{d}^3\mathbf{r}\, \boldsymbol{\nabla} g_{\mu}(\mathbf{r}) \cdot \boldsymbol{\nabla} g^{*}_{\mu'}(\mathbf{r})$
is Hermitian and can therefore be diagonalized by a unitary transformation, $D = U^{\dagger} M U$, with some unitary matrix $U$. This procedure defines a new set of modes $\tilde{g}_{\mu}$ whose gradients satisfy the orthogonality condition of Eq.~\eqref{effective_length}, with the corresponding effective lengths given by the eigenvalues contained in the diagonal matrix~$D$.} The coefficients $\phi_{\mu}(t)$ in Eq.~\eqref{eq : Phi Laplace} serve as the dynamical variables to be quantized, as detailed below. Substituting Eq.~\eqref{eq : Phi Laplace} into Eq.~\eqref{long_pola_electro} yields the general solution for the longitudinal component of the polarization
\begin{equation}
    \label{eq : general P Laplace}
    \mathbf{P}_{\parallel}(\mathbf{r},t) = \varepsilon_0\sum_{\mu} \phi_{\mu}(t) \bm{\nabla}g_{\mu}(\mathbf{r}).
\end{equation}

We now demonstrate that the transverse component of the polarization, $\mathbf{P}_{\perp}$, can be directly inferred from the expression of $\mathbf{P}_{\parallel}$. As previously established, the absence of free volume charges implies that 
\begin{equation}
\bm{\nabla} \cdot \mathbf{P} = 0, \quad \forall \mathbf{r} \notin \Sigma.
\end{equation}
Moreover, under the quasistatic approximation (where time-varying magnetic fields are neglected) the Maxwell-Faraday equation implies $\nabla \times \mathbf{E} = \mathbf{0}$. Both the metallic and dielectric regions are assumed to be linear media, so that the polarization can be expressed as 
\begin{equation}
\label{eqpolachi} 
\mathbf{P} = \varepsilon_0 \chi(\mathbf{r},\omega) \mathbf{E},
\end{equation}
where the dielectric susceptibility is piecewise constant within each region. Specifically, it is given by $\chi(\mathbf{r}, \omega) = \xi(\mathbf{r})\chi_1(\omega) + \left[1- \xi(\mathbf{r}) \right] \chi_{2}$, with $\chi_1(\omega)=\varepsilon_1(\omega) -1$ and $\chi_2=\varepsilon_2 -1$. Under this assumption, one obtains 
\begin{equation}
\bm{\nabla}\times \mathbf{P} = \mathbf{0}, \quad \forall \mathbf{r} \notin \Sigma,
\end{equation}
indicating that the polarization field is irrotational in the bulk. Consequently, the polarization field is both curl-free and divergence-free throughout space, except at the interface $\Sigma$, where material discontinuities occur. Analogously to Eq.~\eqref{eq : general P Laplace}, the transverse polarization is expanded over the set of independent quasistatic modes as $\mathbf{P}_{\perp}=\sum_{\mu} \mathbf{P}_{\perp, \mu}$, such that the total polarization takes the form $\mathbf{P}=\sum_{\mu} \mathbf{P}_{\mu}$, where $\mathbf{P}_{\mu}=\mathbf{P}_{\parallel, \mu}+\mathbf{P}_{\perp, \mu}$ and $\mathbf{P}_{\parallel,\mu}\equiv \varepsilon_0 \phi_{\mu}(t) \bm{\nabla}g_{\mu}(\mathbf{r})$. In Appendix \ref{app_transv_long}, we show that the property of the polarization field being both curl-free and divergence-free implies the relations
\begin{subequations}
\label{eq : P lambda}
\begin{align}
\mathbf{P}_{\perp,\mu} &= - \mathbf{P}_{\parallel, \mu} 
\quad \,\, \mathrm{for} \, \mathbf{r} \in \mathbb{R}^3/V, \\
\label{eq : prop2}\mathbf{P}_{\perp, \mu} &= \lambda_{\mu} \mathbf{P}_{\parallel, \mu} 
\quad \mathrm{for} \, \mathbf{r} \in V.
\end{align}
\end{subequations}

We now exploit these results to evaluate the electronic Lagrangian in Eq.~\eqref{eq:Lagrange_matter2}. The kinetic contribution, which is already expressed as an integral over the metallic region, can be easily handled. By contrast, the integral governing the potential energy extends over the entire space. Nevertheless, the global orthogonality relation between transverse and longitudinal fields, Eq.~\eqref{orth_relation}, can be exploited to write
\begin{equation}
\label{eq : ortho P perp para}
     \int_{\mathbb{R}^3 / V} \mathrm{d}^3 \mathbf{r} \, \mathbf{P}_{\perp} \cdot \mathbf{P}_{\parallel} = -\int_{V} \mathrm{d}^3 \mathbf{r} \, \mathbf{P}_{\perp } \cdot \mathbf{P}_{\parallel}.
\end{equation}
Using that $\mathbf{P}_{\parallel} =-\mathbf{P}_{\perp} $ in the space excluding the metal $\mathbb{R}^3 / V$, we can express the potential energy as two contributions where the  integral runs over the metallic part only
\begin{equation}
    \frac{1}{2\varepsilon_0} \int_{\mathbb{R}^3} \mathrm{d}^3 \mathbf{r} \, \mathbf{P}_{\parallel}^2  = \frac{1}{2\varepsilon_0} \int_{V} \mathrm{d}^3 \mathbf{r} \, \mathbf{P}_{\parallel} \cdot (\mathbf{P}_{\parallel}+\mathbf{P}_{\perp}).
\end{equation}
Using this relation, as well as Eq.~\eqref{eq : prop2}, we can express the matter Lagrangian ~\eqref{eq:Lagrange_matter2} solely as an integral over the metallic region. From the orthogonality equation \eqref{effective_length}, the Lagrangian can be decomposed over the set of harmonic modes $\mu$ as
\begin{equation}
\label{eq : L st final}
     \mathcal{L}_{\mathrm{el}} = \sum_{\mu} \frac{\left( 1 + \lambda_{\mu}\right)}{2\varepsilon_0 \omega_{\mu}^2} \int_{V} \mathrm{d}^3 \mathbf{r} \left(  \dot{\mathbf{P}}_{\parallel,\mu}^2 -\omega^2_{\mu} \mathbf{P}_{\parallel, \mu}^2 \right),
\end{equation}
where we have introduced the characteristic frequencies
\begin{equation}
\omega_{\mu}=\frac{\omega_{\mathrm{p}}}{\sqrt{1 + \lambda_{\mu}}}. 
\label{surf_plasm_freq}
\end{equation}
Equation \eqref{eq : L st final}, involving only the longitudinal component of the polarization, describes a set of harmonic oscillators with eigenfrequencies $\omega_{\mu}$, representing surface plasmons indexed by the quasistatic modes $\mu$. Since $\lambda_{\mu} \geqslant 0$, it follows that $\omega_{\mu} \leqslant \omega_{\mathrm{p}}$, indicating that the surface plasmon frequencies are lower than the bulk plasma frequency due to the presence of the metal-dielectric interface. This behavior can be interpreted as a consequence of the sum rule for plasmon frequencies~\cite{Pitarke, Apell}, which is stated in Appendix \ref{app_transv_long}.

By substituting the surface plasmon frequency $\omega_{\mu}$ given by Eq.~\eqref{surf_plasm_freq} into Eq.~\eqref{eq : Drude}, one obtains a relation connecting $\lambda_{\mu}$ to the value of the dielectric function evaluated at the plasmon frequency, 
\begin{equation}
\label{eq : lambda epsilon}
    \lambda_{\mu} = \varepsilon_{\infty} - 1 - \varepsilon_{1}(\omega_{\mu}).
\end{equation}
Hence, the parameter $\lambda_{\mu}$, which relates the transverse and longitudinal components of the polarization, can be directly inferred from the electrostatic boundary conditions \eqref{eq : BC Phi}. These boundary conditions can be recast, using Eq.~\eqref{eq : lambda epsilon}, into the form
\begin{equation}
\label{eq : lambda phi}
    \left(1 - \varepsilon_{\infty} + \lambda_{\mu}\right) \frac{\partial g_{\mu}(\mathbf{r})}{\partial \mathbf{N}}\Big{|}_1 + \varepsilon_2 \frac{\partial g_{\mu}(\mathbf{r})}{\partial \mathbf{N}}\Big{|}_2 = 0.
\end{equation}

We now return to the electronic Hamiltonian, which, as established in Sec.~\ref{Sec_lagra_Hamil}, depends solely on the total polarization field $\mathbf{P}$ and involves a characteristic frequency corresponding to the bulk plasma frequency $\omega_{\mathrm{p}}$. By combining Eqs.~\eqref{eq : general P Laplace} and \eqref{eq : P lambda}, the total polarization field can be expressed as
\begin{equation}
\mathbf{P}(\mathbf{r},t) = \varepsilon_{0} \xi(\mathbf{r}) \sum_{\mu} \left(1+\lambda_{\mu}\right)  \phi_{\mu} (t) \bm{\nabla}g_{\mu}(\mathbf{r}).
\end{equation}
To obtain its quantized form, we therefore express the dynamical variables as $\phi_{\mu} (t)=\underline{\phi}_{\mu} \mathrm{e}^{-\mathrm{i}\omega_{\mathrm{p}}t} + \underline{\phi}^{*}_{\mu} \mathrm{e}^{+\mathrm{i}\omega_{\mathrm{p}}t}$, where $\underline{\phi}_{\mu}$ denotes the analytic signal composed of spectral components at positive frequencies, and make the substitution
\begin{equation}
 \underline{\phi}_{\mu} \to \frac{L_{\mu}}{1+\lambda_{\mu}} \sqrt{\frac{\hbar \omega_{\mathrm{p}}}{2 \varepsilon_{0}}} B_{\mu}, \qquad \underline{\phi}^{*}_{\mu} \to \frac{L_{\mu}}{1+\lambda_{\mu}} \sqrt{\frac{\hbar \omega_{\mathrm{p}}}{2 \varepsilon_{0}}} B^{\dagger}_{\mu},
\end{equation}
where $B_{\mu}$ and $B^{\dagger}_{\mu}$ are bosonic annihilation and creation operators satisfying the commutation relation $[B_{\mu},B^{\dagger}_{\mu'}]=\delta_{\mu,\mu'}$. The normalization of these operators is chosen such that the electronic Hamiltonian takes the form of a quantum harmonic oscillator
\begin{equation}
\label{H_elec_bos}
H_{\mathrm{el}} = \hbar \omega_{\mathrm{p}} \sum_{\mu} \left(B^{\dagger}_{\mu} B_{\mu}^{\phantom{\dagger}}+\frac 12\right).
\end{equation}
The resulting quantized expression obtained for the polarization field depends explicitly on the effective mode length $L_{\mu}$, which then corresponds to the polarization quantization length, and is given by
\begin{equation}
\label{P_elec_bos}
\mathbf{P}(\mathbf{r}) = \sqrt{\frac{\hbar \omega_{\mathrm{p}}\varepsilon_{0}}{2}} \xi(\mathbf{r}) \sum_{\mu} L_{\mu} B_{\mu}^{\phantom{\dagger}}  \bm{\nabla}g_{\mu}(\mathbf{r})+ \mathrm{H.c.}
\end{equation}
This result \eqref{P_elec_bos} extends the expression derived in Ref.~\cite{Todorov2014} within a bosonization framework based on the long-wavelength response of a three-dimensional electron gas in the random phase approximation. This procedure remains valid as long as nonlocal effects are neglected so that the dielectric function in the metal is of the form of Eq.~\eqref{eq : Drude}. In the present formulation, the geometry of the metallic region is explicitly taken into account. We still assume that this region is sufficiently small so that propagation effects within it can be neglected; this corresponds to the long-wavelength approximation used in the derivation. In Sec.~\ref{sec : SP qs}, we show how this result can be extended to include propagation effects for the case of a planar interface.

\subsection{Free electromagnetic field}
\label{sec_freeph}

Equipped with a quantum description of the electronic degrees of freedom, we now turn to the free quantized EM field within the PZW framework. The vector potential of the free photon field can be expressed, within a quantization volume $V_0 \gg V$, as a plane-wave Fourier expansion \cite{Cohen},
\begin{equation}
\label{eq : A quantum}
     \mathbf{A}\left( \mathbf{r} \right ) = \sum_{\mathbf{k}, \sigma} \sqrt{ \frac{\hbar}{2 V_0 \varepsilon_0 c k}} \bm{\epsilon}_{\mathbf{k},\sigma} a_{\mathbf{k},\sigma} \mathrm{e}^{\mathrm{i} \mathbf{k} \cdot \mathbf{r}}  + \mathrm{H.c.},
\end{equation}
where $a_{\mathbf{k},\sigma}$ ($a_{\mathbf{k}, \sigma}^{\dagger}$) annihilates (creates) a photon of wavevector $\mathbf{k}$ and polarization $\bm{\epsilon}_{\mathbf{k},\sigma}$, with $\sigma=\mathrm{TE},\mathrm{TM}$, where $\mathrm{TE}$ and $\mathrm{TM}$ denote the transverse electric and transverse magnetic polarization directions, respectively. The polarization vectors expressed in terms of the components $\mathbf{k} = \left(k_x, k_y, k_z\right)$ read
\begin{subequations}
\label{eq:TE_TM}
\begin{align}
\bm{\epsilon}_{\mathbf{k},\mathrm{TE}} &= \frac{1}{k_{\parallel}} \left( k_y \mathbf{\hat{x}} - k_x \mathbf{\hat{y}} \right), \\
\bm{\epsilon}_{\mathbf{k},\mathrm{TM}} &= \frac{1}{k k_{\parallel}} \left( k_x k_z \mathbf{\hat{x}} + k_y k_z \mathbf{\hat{y}} - k_{\parallel}^2 \mathbf{\hat{z}} \right),
\end{align}
\end{subequations}
where $\mathbf{\hat{x}}$, $\mathbf{\hat{y}}$ and $\mathbf{\hat{z}}$ denote the three unit vectors directed along the Cartesian axis, $k_{\parallel} = (k_x^2 + k_y^2)^{1/2}$ is the in-plane wavevector, and $k = | \mathbf{k}|$. The creation and annihilation operators satisfy the bosonic commutation relation $[a_{\mathbf{k},\sigma}, a_{\mathbf{k}', \sigma'}^{\dagger}] = \delta_{\mathbf{k}, \mathbf{k}'} \delta_{\sigma,\sigma'}$. In free space, the electric displacement field obeys $\mathbf{D} = \varepsilon_0 \mathbf{E}$. Consequently, the quantized displacement field is given by
\begin{equation}
\label{eq : D quantum}
    \mathbf{D}(\mathbf{r}) = \mathrm{i}\sum_{\mathbf{k}, \sigma} \sqrt{  \frac{\varepsilon_0 \hbar c k}{2 V_0}} \bm{\epsilon}_{\mathbf{k},\sigma} a_{\mathbf{k}, \sigma} \mathrm{e}^{\mathrm{i} \mathbf{k} \cdot \mathbf{r}} + \mathrm{H.c.}
\end{equation}
With the quantized forms of the vector potential and displacement field, the free-photon Hamiltonian \eqref{eq : Hfield}, takes the form
\begin{equation}
    \label{eq : Hph quantum}
    H_{\mathrm{ph}} = \sum_{\mathbf{k}, \sigma} \hbar c k  \left(a_{\mathbf{k}, \sigma}^{\dagger} a_{\mathbf{k}, \sigma}^{\phantom{\dagger}} +\frac 12\right).
\end{equation}

\subsection{Light-matter coupling}
\label{sec_lightmatt}

We can now use the quantized expression of the polarization field \eqref{P_elec_bos} together with the quantized free-space displacement field \eqref{eq : D quantum} to derive the light-matter interaction term \eqref{eq : Hint}. This provides 
\begin{equation}
    \label{eq : Hint quantum}
    H_{\mathrm{int}} = -\mathrm{i} \sum_{\mu,\mathbf{k},\sigma} \hbar B_{\mu}^{\phantom{\dagger}}  \left( C_{\mu,\mathbf{k},\sigma} a_{\mathbf{k}, \sigma}^{\phantom{\dagger}} - \tilde{C}_{\mu,\mathbf{k},\sigma} a^{\dagger}_{\mathbf{k}, \sigma} \right) + \mathrm{H.c.},
\end{equation}
where the coupling strengths read
\begin{subequations}
    \label{couplin_strength}    
    \begin{align}
   C_{\mu,\mathbf{k},\sigma} &= \sqrt{\frac{\omega_{\mathrm{p}} c k}{4 V_{0}}} L_{\mu} \int_{V} \mathrm{d}^3 \mathbf{r} \, \bm{\nabla}g_{\mu}(\mathbf{r}) \cdot \bm{\epsilon}_{\mathbf{k},\sigma} \,\mathrm{e}^{+\mathrm{i} \mathbf{k} \cdot \mathbf{r}}, \\
     \tilde{C}_{\mu,\mathbf{k},\sigma} &= \displaystyle \sqrt{\frac{\omega_{\mathrm{p}} c k}{4 V_{0}}} L_{\mu} \int_{V} \mathrm{d}^3 \mathbf{r} \, \bm{\nabla}g_{\mu}(\mathbf{r}) \cdot \bm{\epsilon}_{\mathbf{k},\sigma}\, \mathrm{e}^{-\mathrm{i} \mathbf{k} \cdot \mathbf{r}}.
    \end{align}
\end{subequations}

By collecting the contributions to the total Hamiltonian $H_{\mathrm{el}} + H_{\mathrm{ph}} + H_{\mathrm{int}}$ from Eqs.~\eqref{H_elec_bos}, \eqref{eq : Hph quantum}, and \eqref{eq : Hint quantum}, it is instructive to observe that, within the PZW framework, the fundamental matter oscillator coupled to the EM field is the bulk plasmon at frequency $\omega_{\mathrm{p}}$. The influence of the metallic boundaries is encoded in the coefficients \eqref{couplin_strength}, which characterize the coupling of this mode to the radiative continuum. In the PZW representation, it is therefore the interaction with radiation modes that renormalizes the bulk plasma frequency and gives rise to the manifold of surface plasmon resonances. Indeed, while electrons couple to light through the transverse component of the polarization field $\mathbf{P}_{\perp}$ in the PZW picture, the presence of interfaces induces correlations between the longitudinal and transverse components of the polarization. As a consequence, the effects of the light-matter interaction with purely transverse photons can be expressed in terms of the electrostatic energy distribution of the oscillating electrons.

More precisely, owing to the proportionality between the transverse and longitudinal components of the polarization [Eq.~\eqref{eq : P lambda}], the full polarization field can be determined by solving Laplace's equation subject to the boundary condition in Eq.~\eqref{eq : lambda phi}. This framework provides a rigorous foundation for deriving the electronic Lagrangian in the quasistatic limit [Eq.~\eqref{eq : L st final}], which describes surface plasmon modes arising from collective oscillations of the electron plasma near the interface. These modes possess a frequency $\omega_{\mu}$ [Eq.~\eqref{surf_plasm_freq}], redshifted relative to the bulk plasma frequency $\omega_{\mathrm{p}}$, while the electronic Hamiltonian \eqref{eq : Hmat 1} accounts for bulk electron-density oscillations at the characteristic frequency $\omega_{\mathrm{p}}$. Within this formalism, we have established a quantized representation of the polarization field incorporating an effective mode length, which, together with the quantized free-space displacement field \eqref{eq : D quantum}, forms the basis for deriving the light-matter coupling Hamiltonian. In Sec.~\ref{Sec_cases}, we will employ this generic light-matter Hamiltonian, given by Eqs.~\eqref{H_elec_bos}, \eqref{eq : Hph quantum}, and \eqref{eq : Hint quantum}, and specialize it to two representative geometries.

\section{Case study based on the general framework}
\label{Sec_cases}

In this section, we apply the general framework developed in Sec.~\ref{sec_electro} to two representative geometries: a spherical nanoparticle, discussed in Sec.~\ref{sec : LSP qs}, and a planar metal-dielectric interface, examined in Sec.~\ref{sec : SP qs}.

\subsection{Localized surface plasmon polaritons of a spherical nanoparticle}
\label{sec : LSP qs}

\begin{figure}
\includegraphics[scale=0.45]{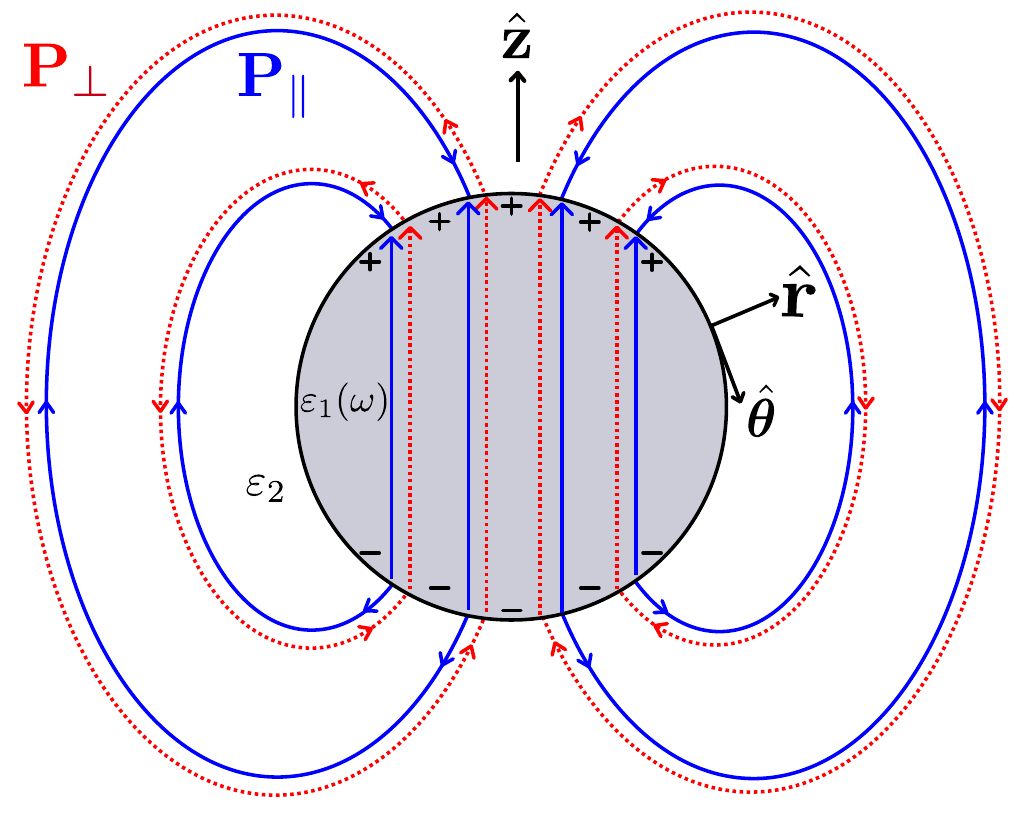}
\caption{Spherical metallic nanoparticle of radius $a$ with a dielectric function $\varepsilon_1(\omega)$ [cf.\ Eq.~\eqref{eq : Drude}], surrounded by a dielectric medium of dielectric constant $\varepsilon_2$. Illustrations of the longitudinal (solid blue lines) and transverse (red dotted lines) polarization fields for the fundamental localized plasmon mode ($l = 1$) of the nanoparticle are represented. The longitudinal and transverse polarizations are constant inside the sphere and oriented along the unit vector $\hat{\mathbf{z}}$. The field lines of $\mathbf{P}_{\perp}$ form closed loops, whereas those of $\mathbf{P}_{\parallel}$ originate at the negative charges and terminate at the positive charges.} 
\label{fig : sphere}
\end{figure}

We consider a homogeneous spherical metallic nanoparticle of radius $a$ (volume $V = 4\pi a^{3} / 3$) embedded in a dielectric medium with dielectric constant $\varepsilon_{2}$, as illustrated in Fig.~\ref{fig : sphere}. Our objective is to determine the frequency and radiative damping rate of an LSP mode coupled to the continuum of free-space photons. The analysis is mostly restricted to the lowest-order dipolar mode; however, we also provide general results that include higher-order modes wherever possible. In all cases, we consider the regime where the nanoparticle radius is much smaller than the photon wavelength, such that propagation effects remain small.

The general solution to the Laplace equation \eqref{eq : Phi Laplace} in spherical coordinates $(r, \theta, \varphi)$ is given by $\phi(\mathbf{r},t) =\sum_{l=0}^{\infty} \phi_{l}(t)  g_{l}(\mathbf{r})$~\cite{Jackson1998}, with
\begin{equation}
    \label{eq : Laplace sphere}
   g_{l} (\mathbf{r})= \sqrt{\frac{2 l +3}{a^{2l+3}}}
   \begin{cases}
r^l Y_l^0 (\theta), &  r \leqslant a,\\[.1cm]
\displaystyle \frac{a^{2l+1}}{r^{l+1}} Y_l^0 (\theta), &  r > a,
\end{cases}    
\end{equation}
where $l$ is an integer separation index, and $Y_{l}^{0}(\theta)$ denotes the spherical harmonics. Note that here we have explicitly broken the symmetry of the system, considering electronic oscillations along a particular axis which is the $z$ axis in Fig.~\ref{fig : sphere}.   As shown in the previous section, the constants $\lambda_{l}$ must be determined to obtain the LSP frequencies $\omega_{l}$ in the quasistatic limit. These frequencies correspond to the collective multipolar oscillations of the conduction electron density near the nanoparticle surface. By substituting Eq.~\eqref{eq : Laplace sphere} into Eq.~\eqref{eq : lambda phi} with $\mathbf{N} \equiv \mathbf{\hat{r}}$, and using the unit vectors $\mathbf{\hat{r}}$ and $\bm{\hat{\theta}}$ defined in Fig.~\ref{fig : sphere}, we find $\lambda_{l} = \varepsilon_{\infty} - 1 + \varepsilon_2 (l+1)/l$. Using Eq.~\eqref{surf_plasm_freq}, the surface plasmon frequency becomes
\begin{equation}
\label{eq : omega l}
\omega_{l} = \omega_{\mathrm{p}}\sqrt{\frac{l}{l\left( \varepsilon_{\infty} + \varepsilon_2 \right) + \varepsilon_2}},
\end{equation}
which coincides with the result of Mie theory~\cite{Mie,Bohren, Kreibig}.
Moreover, the gradients of the solutions \eqref{eq : Laplace sphere} of the Laplace equation verify the orthogonality relation stated in Eq.~\eqref{effective_length}. From the latter, we obtain expressions for the quantization lengths $L_{l}$. They are given by  $L_l = a/ \sqrt{l\left( 2l + 3 \right)}$ and decrease for increasing values of $l$, meaning that the effective volume of plasma oscillations decreases with the order $l$.

We focus on the dipolar mode ($l = 1$) of a nanosphere in vacuum with $\varepsilon_{2} = 1$ and $\varepsilon_{\infty} = 1$. In this case, $\lambda_{1} = 2$ and $\omega_{1} \equiv \omega_{\mathrm{M}} = \omega_{\mathrm{p}} / \sqrt{3}$, corresponding to the Mie frequency. For $l = 1$, the effective mode length reads $L_{1} = a / \sqrt{5}$. The polarization field \eqref{P_elec_bos} then takes the form
\begin{equation}
\label{pola_quant_sphere}
\mathbf{P}(\mathbf{r}) = \sqrt{\frac{\hbar \omega_{\mathrm{p}}\varepsilon_{0}}{2 V}} \Theta (a - r) \left(B+ B^{\dagger} \right)\mathbf{\hat{z}},
\end{equation}
where $B \equiv B_{1}$ and $B^{\dagger} \equiv B_{1}^{\dagger}$, $\Theta$ is the Heaviside function, and $\mathbf{\hat{z}} = \cos\theta \, \mathbf{\hat{r}} - \sin\theta \, \bm{\hat{\theta}}$.

The field lines associated with the transverse and longitudinal components of the polarization are illustrated in Fig.~\ref{fig : sphere}, elucidating the physical content of Eqs.~\eqref{eq : P lambda}. The longitudinal component $\mathbf{P}_{\parallel}$ and the transverse component $\mathbf{P}_{\perp}$ exhibit distinct topological structures. The field $\mathbf{P}_{\perp}$ is solenoidal throughout space, implying that its field lines form closed loops and cross the metal boundary without interruption, although they can be broken. By contrast, the field lines of $\mathbf{P}_{\parallel}$ are discontinuous at the metal surface: They originate from negative charges and terminate on positive charges. Outside the particle, each field line of $\mathbf{P}_{\perp}$ is accompanied by a corresponding field line of $\mathbf{P}_{\parallel}$ oriented in the opposite direction, such that the total polarization field vanishes outside the metal. Inside the metal, both components of the polarization field are uniform and aligned along the $z$ axis; however, the magnitude of $\mathbf{P}_{\parallel}$ is half that of $\mathbf{P}_{\perp}$. This reflects the partial screening of electric field lines within the metal due to dynamical charge oscillations. Note that outside the sphere the longitudinal component $\mathbf{P}_{\parallel}$ corresponds, up to a multiplicative constant, to the electric field of a dipolar charge distribution. This description remains valid for higher-order multipolar modes ($l > 1$) \cite{Jackson1998}.

Combining Eqs.~\eqref{H_elec_bos}, \eqref{eq : Hph quantum}, and \eqref{eq : Hint quantum} for the lowest-order dipolar mode, which couples only to TM photon modes, the Hamiltonian reads
\begin{align}
\label{full_H_sphere}
H =&\; \hbar \omega_{\mathrm{p}} B^{\dagger} B
+ \sum_{\mathbf{k}} \hbar c k \, a_{\mathbf{k}}^{\dagger} a_{\mathbf{k}}^{\phantom{\dagger}} \nonumber \\
&+ \mathrm{i} \sum_{\mathbf{k}} \hbar C_{\mathbf{k}}
\left( B + B^{\dagger} \right)
\left( a^{\phantom{\dagger}}_{\mathbf{k}} - a^{\dagger}_{-\mathbf{k}} \right),
\end{align}
where $ a^{\phantom{\dagger}}_{\mathbf{k}} \equiv a^{\phantom{\dagger}}_{\mathbf{k},\mathrm{TM}}$, and the light-matter coupling strength is
\begin{equation}
C_{\mathbf{k}} = \sqrt{\frac{\omega_{\mathrm{p}} c k}{4 V_{0} V}} \left( \frac{k_{\parallel}}{k} \right) f(k),
\end{equation}
with the structure factor 
\begin{equation}
f(k)=\int_{V} \mathrm{d}^3\mathbf{r}  \,  \mathrm{e}^{\mathrm{i} \mathbf{k} \cdot \mathbf{r}} =\frac{3 V}{(ka)^2} \left[ \frac{\sin (ka)}{ka} - \cos (ka)\right],
\end{equation}
that is essentially the Fourier transform image of a spherical shape. Interestingly, in the minimal coupling representation, such a geometric factor does not appear explicitly in the Hamiltonian as it is already included in the fundamental matter oscillator frequency \cite{Allard}.

The Hamiltonian~\eqref{full_H_sphere} describes the dipolar plasmonic mode renormalized by its interaction with the photonic continuum, which induces both a frequency shift and a radiative decay rate due to photon emission. These quantities can be obtained, for instance, through the Fano formalism~\cite{Fano1961,DeLiberato2022,Feist2025}. 

Here, we adopt the Green's function formalism, a powerful and widely used technique in condensed-matter physics~\cite{Bruus2004}. Our treatment is simplified by the fact that the Hamiltonian~\eqref{full_H_sphere} effectively describes a single harmonic oscillator mode that is nonuniformly coupled to a bath of harmonic oscillators. The time-ordered plasmon Green's function at zero temperature is defined as
\begin{equation}
    \label{eq : Green plasmon}
    \mathcal{G}(t-t') = - \mathrm{i} \langle \mathcal{T} X(t)X(t') \rangle,
\end{equation}
where $X =B + B^{\dagger}$, and $\mathcal{T}$ denotes the time-ordering operator, defined as $\mathcal{T} X(t)X(t') = \Theta(t-t') X(t) X(t') + \Theta(t'-t) X(t')X(t)$, while $\langle \dots \rangle$ denotes the expectation value in the ground state. By differentiating the Green's function twice with respect to time and using Heisenberg's equation of motion, one obtains
\begin{align}
    \label{eq : EOM}
    \frac{\partial^2 \mathcal{G}(t - t')}{\partial t^2}
    =&\; \frac{\delta(t - t')}{\hbar} \left\langle \left[ \left[ H, X(t) \right], X(t) \right] \right\rangle \nonumber \\
    & + \frac{\mathrm{i}}{\hbar^2} \left\langle \mathcal{T} \left[ H, \left[ H, X(t) \right] \right] X(t') \right\rangle.
\end{align}
This leads, using Eq.~\eqref{full_H_sphere} and the bosonic commutation relation of the $B$ operators, to
\begin{align}
\label{eq : EOM 1}
\frac{\partial^2 \mathcal{G}(t - t')}{\partial t^2}
=& - 2 \omega_{\mathrm{p}} \delta(t - t') 
- \omega_{\mathrm{p}}^2 \mathcal{G}(t - t') \nonumber \\
& - 2\mathrm{i} \omega_{\mathrm{p}} \sum_{\mathbf{k}} C_{\mathbf{k}} \mathcal{G}_{\mathbf{k}}(t - t'),
\end{align}
where we have introduced the mixed plasmon-photon Green's function $\mathcal{G}_{\mathbf{k}}(t-t') = - \mathrm{i} \langle \mathcal{T} D_{\mathbf{k}}(t)X(t') \rangle$, with $D_{\mathbf{k}} = a^{\phantom{\dagger}}_{\mathbf{k}} - a^{\dagger}_{\mathbf{k}}$. A second equation of motion can be derived for the mixed Green's function:
\begin{equation}
    \label{eq : EOM 2}
    \frac{\partial^2 \mathcal{G}_{\mathbf{k}}(t-t')}{\partial t^2} =   - c^2 k^2 \mathcal{G}_{\mathbf{k}}(t-t')  + 2\mathrm{i} ck C_{\mathbf{k}} \mathcal{G}(t-t').
\end{equation}
Introducing the Fourier transforms $\mathcal{G}(\Omega)$ and $\mathcal{G}_{\mathbf{k}} (\Omega)$ through $\mathcal{G}(t-t')=\int \! \frac{\mathrm{d}\Omega}{2\pi}\, \mathrm{e}^{-\mathrm{i} \Omega (t-t')} \mathcal{G}(\Omega)$, and a similar definition for $\mathcal{G}_{\mathbf{k}} (\Omega)$, Eq.~\eqref{eq : EOM 2} yields, in the frequency domain,
\begin{equation}
    \label{eq : correlation GF}
    \mathcal{G}_{\mathbf{k}}(\Omega) = \frac{2 \mathrm{i} ck C_{\mathbf{k}}}{c^2k^2 - \Omega^2} \mathcal{G}(\Omega).
\end{equation}
Substituting Eq.~\eqref{eq : correlation GF} into Eq.~\eqref{eq : EOM 1} written in the frequency domain, the retarded (causal) plasmon Green's function, defined in real time as $\mathcal{G}^{\mathrm{R}} (t - t') = -\mathrm{i} \,\theta(t - t') \,\langle [X(t), X(t')] \rangle$, can be expressed in the Dyson form as
\begin{equation}
    \label{eq : Dyson plasmon}
    {\mathcal{G}^{\mathrm{R}}}(\Omega) = \frac{1}{\mathcal{G}^{-1}_{0}(\Omega) - \Sigma^{\mathrm{R}}(\Omega)},
\end{equation}
where $\mathcal{G}_{0}(\Omega) = 2\omega_{\mathrm{p}}/(\Omega^2 - \omega_{\mathrm{p}}^2)$ denotes the bare plasmon Green's function, and 
\begin{equation}
    \label{eq : Self energy}
    \Sigma^{\mathrm{R}}(\Omega) =  \sum_{\mathbf{k}} \frac{2ck C^2_{\mathbf{k}}}{\left( \Omega + \mathrm{i} 0^{+}\right)^2 - c^2k^2} 
\end{equation}
is the exact retarded self-energy arising from the plasmon-photon coupling. The retarded character of the self-energy has been incorporated by replacing $\Omega$ with $\Omega + \mathrm{i}0^{+}$ in the denominator. The explicit calculation of this self-energy is presented in Appendix \ref{app_met_nano} within the quasistatic (long-wavelength) limit $\Omega a / c \ll 1$. For noble-metal nanoparticles with $a \sim \unit[10]{nm}$ and $\hbar \Omega \sim \hbar \omega_{\mathrm{p}} \sim \unit[10]{eV}$, this approximation is well justified. The frequency renormalization and decay rate of the plasmonic mode can be extracted from the plasmon spectral function, defined as 
$A(\Omega) \equiv -2\, \mathrm{Im}\left\{ \mathcal{G}^{\mathrm{R}}(\Omega)\right\}$. As shown in Appendix \ref{app_met_nano}, this function exhibits a peak at $\Omega = \omega_{\mathrm{M}} + \delta$, where the leading-order radiative frequency shift in the small parameter $\omega_{\mathrm{M}} a / c$, which quantifies deviations from the quasistatic regime, is given by
\begin{equation}
\label{eq : lamb shift}
\delta = - \frac{2\omega_{\mathrm{M}}}{5} \left( \frac{\omega_{\mathrm{M}} a}{c} \right)^2.
\end{equation}
This expression for the radiative redshift exactly matches the classical Drude-Lorentz result at leading-order in $\omega_{\mathrm{M}} a / c$~\cite{Allard,verde2025}. Furthermore, Eq.~\eqref{eq : lamb shift} is consistent with the result obtained in Ref.~\cite{Allard} using second-order perturbation theory in the minimal-coupling representation. In that formulation, an ultraviolet cutoff must be introduced to regularize the frequency integrals governing the radiative redshift. Notably, no such cutoff is required when the calculation is performed in the PZW representation.

The width of the spectral function $A(\Omega)$ gives the rate at which LSPs decay in the photon continuum. It is expressed, at leading order in $\omega_{\mathrm{M}} a / c$, as
\begin{equation}
\label{eq : result gamma}
\Gamma =  \frac{2\omega_{\mathrm{M}}}{3} \left(\frac{\omega_{\mathrm{M}}a}{c} \right)^3.
\end{equation}
In the quasistatic limit, $\omega_{\mathrm{M}} a / c \ll 1$, Eq.~\eqref{eq : result gamma} coincides with the classical result for the radiation of an oscillating electric dipole~\cite{Jackson1998}, as well as with the perturbative result of Ref.~\cite{Allard} obtained within the minimal-coupling representation.

At this point, it is worth emphasizing the fundamental distinction between the minimal-coupling and PZW formulations of QED for metal-dielectric interfaces. In the minimal-coupling approach, the electronic Hamiltonian describes an oscillator at the LSP frequency. Radiative corrections, namely the redshift $\delta$ and the damping rate $\Gamma$, arise directly from the light-matter interaction, and the $\mathbf{A}^{2}$ term can be shown not to contribute. By contrast, in the PZW representation, the electronic Hamiltonian corresponds to an oscillator at the bulk plasma frequency. Consequently, the light-matter coupling produces a redshift of order unity in addition to the small correction $\delta$. This behavior highlights the intrinsically nonperturbative nature of light-matter interactions in metallic nanoparticle systems within the PZW framework. In Sec.~\ref{sec : SP qs}, we show that the same conclusion holds for a planar metal-dielectric interface.

We now demonstrate that the plasmon-photon coupling in the PZW representation is not only intrinsically nonperturbative but also ultrastrong. For simplicity, we focus on the quasistatic limit $\omega_{\mathrm{M}} a / c \to 0$, in which $\Gamma \to 0$ according to Eq.~\eqref{eq : result gamma}. In this regime, radiative losses can be neglected, allowing the Hamiltonian \eqref{full_H_sphere} to be diagonalized using the Hopfield-Bogoliubov method~\cite{Hopfield}. The Hamiltonian is written as $H = \hbar \Omega p^{\dagger} p$, where the LSP polariton annihilation operator is 
\begin{equation}
    \label{eq : pol sphere}
    p = w B + xB^{\dagger} + \sum_{\mathbf{k}} \left( y_{\mathbf{k}} a_{\mathbf{k}}^{\phantom{\dagger}} + z_{\mathbf{k}} a_{\mathbf{k}}^{\dagger}\right),
\end{equation}
and the Hopfield coefficients satisfy
\begin{equation}
    \label{eq : normalization sphere}
    \vert w\vert^2 -\vert x \vert^2 + \sum_{\mathbf{k}} \left( \vert y_{\mathbf{k}}\vert^2 - \vert z_{\mathbf{k}}\vert^2\right) = 1.
\end{equation}
Solving the Heisenberg equation of motion $[p, H] = \hbar \Omega p$ yields the coupled system of equations
\begin{subequations}
\label{eq : Hopfield sphere}
\begin{align}
w \left( \omega_{\mathrm{p}} - \Omega \right)
&= +\mathrm{i} \sum_{\mathbf{k}} C_{\mathbf{k}} \left( y_{\mathbf{k}} + z_{\mathbf{k}} \right), \\
x \left( \omega_{\mathrm{p}} + \Omega \right)
&= -\mathrm{i} \sum_{\mathbf{k}} C_{\mathbf{k}} \left( y_{\mathbf{k}} + z_{\mathbf{k}} \right), \\
y_{\mathbf{k}} \left( c k - \Omega \right)
&= \mathrm{i} C_{\mathbf{k}} \left( x - w \right), \\
z_{\mathbf{k}} \left( c k + \Omega\right)
&= \mathrm{i} C_{\mathbf{k}} \left( x - w \right),
\end{align}
\end{subequations}
which, after eliminating the Hopfield coefficients, yields the eigenfrequency equation
\begin{equation}
\label{eq : eigenvalue LSP}
    \Omega^2 = \omega_{\mathrm{p}}^2 - \sum_{\mathbf{k}} \frac{4 c k \omega_{\mathrm{p}} C_{\mathbf{k}}^2}{c^2 k^2 - \Omega^2}.
\end{equation}
In the quasistatic limit $\omega_{\mathrm{M}} a / c \to 0$, the LSP polariton eigenfrequency approaches $\Omega = \omega_{\mathrm{M}}$, consistent with Eq.~\eqref{eq : lamb shift}, which predicts a vanishing frequency shift. A convenient quantity to characterize the LSP polariton is its total electronic weight, defined via the normalization condition in Eq.~\eqref{eq : normalization sphere} as $\eta_{\mathrm{el}} = |w|^{2} - |x|^{2}$. Combining Eqs.~\eqref{eq : normalization sphere} and \eqref{eq : Hopfield sphere} yields
\begin{equation}
    \label{eq : electronic weight LSP}
    \eta_{\mathrm{el}} = 
    \left[{1 + \sum_{\mathbf{k}} 
    \frac{4 c k \omega_{\mathrm{p}} C^{2}_{\mathbf{k}}}
    {\left( c^{2} k^{2} - \Omega^{2} \right)^{2}} }\right]^{-1}.
\end{equation}
Replacing the summation over $\mathbf{k}$ with an integral, performing the angular integration, and taking the limit $\Omega a / c \to 0$ directly yields $\eta_{\mathrm{el}} = 1$, indicating that the LSP polariton in the PZW representation is purely electronic in character. We emphasize that the electronic and photonic content of the polariton cannot be accessed within the Green's function formalism. In contrast, these quantities arise naturally within a Hopfield-Bogoliubov approach formulated for normal modes, i.e., in the limit of vanishing losses.

Because the Hamiltonian \eqref{full_H_sphere} contains counter-rotating terms of the form $B a^{\phantom{\dagger}}_{\mathbf{k}}$ and $B^{\dagger} a^{\dagger}_{\mathbf{k}}$, the ground state of the coupled system is no longer the bare vacuum but instead the polaritonic vacuum, defined by $p \vert G \rangle = 0$. Since the polariton annihilation operator is a linear combination of both creation and annihilation operators for plasmons and photons, the ground state $\vert G \rangle$ takes the form of a squeezed state~\cite{Ciuti2005}. 

In particular, the number of \textit{bulk} plasmons in the ground state is finite and is given by the expectation value $\langle G \vert B^{\dagger} B \vert G \rangle$. Since $\eta_{\mathrm{el}} = 1$, Eq.~\eqref{eq : pol sphere} implies that $B = w^{*} p - x\, p^{\dagger}$. Consequently, the ground-state plasmon population takes the form $\langle G \vert B^{\dagger} B \vert G \rangle = |x|^{2}$. 
Using the relations between the Hopfield coefficients obtained by combining Eqs.~\eqref{eq : normalization sphere} and \eqref{eq : Hopfield sphere}, the bulk plasmon population in the ground state becomes
\begin{equation}
    \label{eq : pop plasmons sphere}
    \langle G \vert B^{\dagger} B \vert G \rangle
    = \frac{\left(\omega_{\mathrm{p}}-\Omega \right)^{2}}{4\, \omega_{\mathrm{p}}\, \Omega}
\end{equation}
in the quasistatic limit. 
In the case where 
$\varepsilon_\infty=\varepsilon_2=1$,  $\Omega\simeq\omega_\mathrm{M}=\omega_\mathrm{p}/\sqrt{3}$, we have
$\langle G \vert B^{\dagger} B \vert G \rangle=1/\sqrt{3}-1/2\simeq 7.7\times 10^{-2}$.
The bulk plasmon population in the polaritonic ground state is virtual and should be interpreted as a manifestation of anomalous quantum fluctuations due to ultrastrong coupling. While this population is finite, it remains nonmacroscopic. This property is ensured by the presence of the $\mathbf{P}^2$ term in our model, which guarantees the stability of the ground state, in contrast to Dicke-type models~\cite{EmaryBrandes2003}. 

Equations \eqref{eq : eigenvalue LSP}--\eqref{eq : pop plasmons sphere} are formally analogous to the corresponding expressions in the Hopfield model~\cite{Hopfield}, although the physical situation considered here is somewhat different. In our case, the system involves a mode-to-continuum coupling rather than the mode-to-mode coupling characteristic of the Hopfield model. This difference originates from the fact that the electronic and photonic degrees of freedom are defined on domains with distinct symmetry properties, which prevents a decomposition into independent momentum sectors as in the Hopfield model. Nevertheless, guided by the formal analogy, we extend the concept of ultrastrong coupling to this mode-to-continuum scenario by adopting the plasmon population as the relevant figure of merit, even though no well-defined coupling ratio, the standard figure of merit for ultrastrong coupling, can be introduced in this case.

\begin{figure}
\includegraphics[width=\columnwidth]{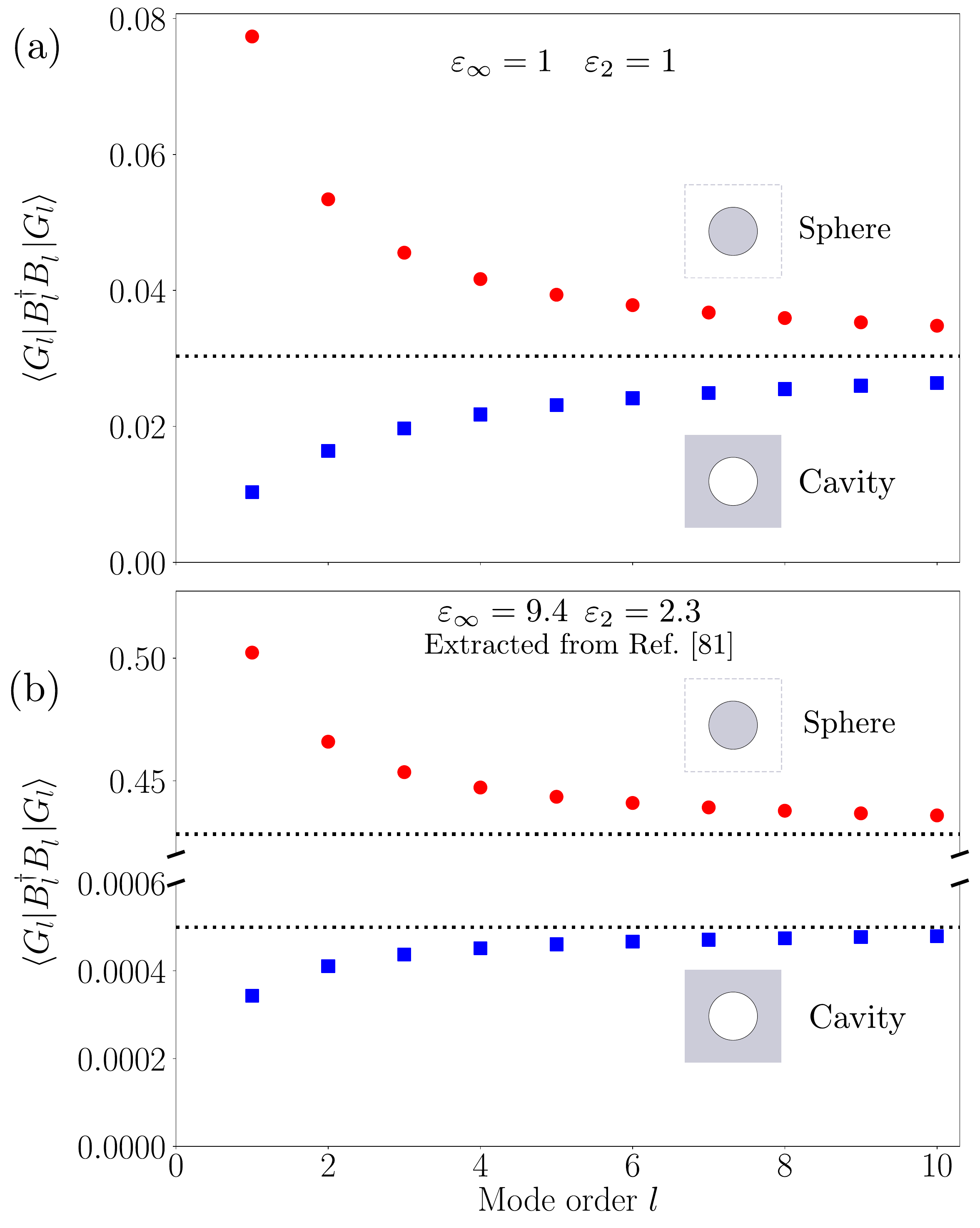}
\caption{(a) Ground-state population of bulk plasmons for a metallic sphere (red dots) and its complementary spherical cavity (blue squares) as a function of the mode order $l$ for $\varepsilon_2 = 1$ and $\varepsilon_{\infty} = 1$. The dotted line corresponds to the asymptotic limit of a flat interface. (b) Same quantities as in panel (a) for $\varepsilon_2 = 2.3$ and $\varepsilon_{\infty} = 9.4$ extracted from Ref.~\cite{Berciaud2005}. The dotted lines are the asymptotic limits.} 
\label{fig : GSPop}
\end{figure}

The expression in Eq.~\eqref{eq : pop plasmons sphere} is in fact quite general and can be extended to higher-order modes, for which the eigenfrequencies are given by $\omega_l = \omega_{\mathrm{p}}\sqrt{l/(2l+1)}$, as well as to the complementary geometry of a spherical cavity embedded in a metal, characterized by $\tilde{\omega}_l = \omega_{\mathrm{p}}\sqrt{(l+1)/(2l+1)}$ (see Appendix~\ref{app_transv_long}). The corresponding ground-state plasmon populations are shown in Fig.~\ref{fig : GSPop}(a). The maximum population occurs for the lowest-order dipolar ($l=1$) mode of a spherical nanoparticle, whereas the minimum is found for the lowest-order mode of a spherical cavity. In the latter case, the quantum ground state is expected to be closest to that of a bulk plasma. For larger angular-momentum indices $l$, both geometries converge toward a common asymptotic value, which, as discussed below, corresponds to the planar-interface limit. 

Finally, we note that although the nonzero plasmon populations displayed in Fig.~\ref{fig : GSPop}(a) remain relatively small, they can be substantially enhanced by considering dielectric constants $\varepsilon_{\infty}$ and $\varepsilon_{2}$ larger than unity. Indeed, as can be seen from Eq.~\eqref{eq : omega l}, the LSP resonance frequency depends on the values of these dielectric constants. Consequently, such a  dependence carries over to the plasmon populations in the ground state \eqref{eq : pop plasmons sphere} since $\Omega\simeq\omega_l$.
For instance, the experiments of Ref.~\cite{Berciaud2005} on single gold nanoparticles with a radius $a = \unit[2.5]{nm}$ and $\hbar\omega_{\mathrm{p}} = \unit[9.0]{eV}$ embedded in a medium of dielectric constant $\varepsilon_2 = 2.3$, reported a dipolar LSP resonance frequency $\hbar\omega_\mathrm{M} \simeq \unit[2.4]{eV}$. From these parameters, $\varepsilon_{\infty} = 9.4$ can be inferred using Eq.~\eqref{eq : omega l} for $l=1$. The corresponding plasmon populations as a function of the mode order $l$ are displayed in Fig.~\ref{fig : GSPop}(b). In particular, we find that the predicted bulk plasmon population in the ground state for the dipolar mode can reach a large value $\sim 0.5$. Interestingly, the asymptotic limits of the populations for the sphere and the cavity no longer coincide when $\varepsilon_2 \neq 1$ and $\varepsilon_{\infty} \neq 1$. This can be understood by noting that, for large values of $l$, $\omega_l \rightarrow \omega_{\mathrm{p}}\sqrt{1/(\varepsilon_{\infty} + \varepsilon_2})$, whereas, due to the sum rule for plasma frequencies \eqref{eq : sum rule freq}, $\tilde{\omega}_l \rightarrow \omega_{\mathrm{p}} \sqrt{1 - 1/(\varepsilon_{\infty} + \varepsilon_2})$. These asymptotic frequencies differ unless $\varepsilon_{\infty} + \varepsilon_2 = 2$. Consequently, the sum rule explains why the increase in the populations for the sphere between Fig.~\ref{fig : GSPop}(a) and Fig.~\ref{fig : GSPop}(b) is accompanied by a decrease in the populations for the cavity.

Several strategies may be envisioned to probe the anomalous ground-state population. A first possibility, originally proposed in Ref.~\cite{Ciuti2005}, relies on nonadiabatic modulations of the system parameters on timescales shorter than the optical cycle, thereby converting virtual excitations into real emitted photons~\cite{Ciuti2005,Forn-Diaz2019,kockum2019ultrastrong}. In the present context, one could in principle modulate the plasma frequency. While this is impractical in metallic nanoparticles due to their large electronic density, semiconductor plasmonic nanostructures provide a promising alternative, as they support lower-frequency plasmonic resonances \cite{Guo2025} and allow ultrafast carrier-density modulation~\cite{Taliercio_2012,Gunter2009}. Another possible route consists in modulating the dielectric environment surrounding the nanoparticle on deep-subcycle timescales. In this respect, order-unity refractive-index variations have recently been demonstrated in transparent conducting oxides operating in the epsilon-near-zero regime~\cite{Tirole2023,Lustig2023,Segal2026}. A complementary approach would rely on the direct sampling of vacuum fluctuations using electro-optic sampling techniques, as demonstrated in Refs.~\cite{Riek2015,BeneaChelmus2019}. Another interesting possibility is provided by dynamical Coulomb blockade~\cite{Iqbal_2024} in scanning tunneling microscopy geometries involving metallic nanoparticles~\cite{Brun_2012,Senkpiel_2020}, where the tunneling current is influenced by the quantum fluctuations of the nanoparticle impedance. In currently available experiments, however, the effective impedance is dominated by a resistance-capacitance response originating from nonradiative losses, whereas probing anomalous populations requires an inductive-capacitive regime. Here again, semiconductor plasmonic nanostructures may constitute a more suitable platform. Finally, another possible route would be to strongly couple a two-level emitter to the nanoparticle near field. In such a configuration, the anomalous photon population could affect the quantum statistics of spontaneous emission or nonlinear optical emission under strong driving. A quantitative theoretical analysis of this regime remains an open problem.

\subsection{Propagating surface plasmon polaritons at a planar interface}
\label{sec : SP qs}

\begin{figure}
    \centering
    \includegraphics[scale=0.4]{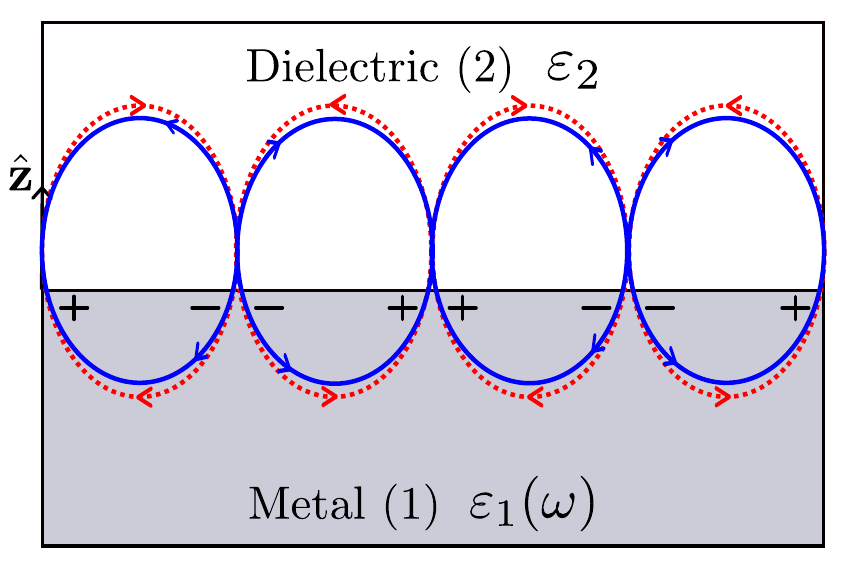}
    \caption{Planar interface separating a metal with dielectric function $\varepsilon_1(\omega)$ for $z<0$ and a dielectric medium of constant $\varepsilon_2$ for $z>0$. The field lines of longitudinal polarization (blue, solid) and transverse polarization (red, dotted) are shown. As displayed in Fig.~\ref{fig : sphere}, the field lines of $\mathbf{P}_{\perp}$ are closed loops and those of $\mathbf{P}_{\parallel}$ start at the negative charges and end at the positive charges. }
    \label{fig : interface plane}
\end{figure}

As a second example, we now examine a planar interface between a semi-infinite metal ($z<0$) and a semi-infinite dielectric ($z>0$) along the $z$ axis, as illustrated in Fig.~\ref{fig : interface plane}. As in the previous section, our goal is to analyze the eigenmodes of such a system within our framework, namely PSP polaritons propagating along the interface at $z=0$. To this end, we first rely on the results of Sec.~\ref{sec_quasilagrang} and characterize PSPs in the quasistatic regime.  

The general solution of the Laplace equation in this geometry reads $\phi(\mathbf{r},t) =\sum_{\mathbf{k}_{\parallel}} \phi_{\mathbf{k}_{\parallel}}(t)  g_{\mathbf{k}_{\parallel}}(\mathbf{r})$, with
\begin{equation}
\label{eq : Phi Plan}
   g_{\mathbf{k}_{\parallel}}(\mathbf{r}) = \sqrt{\frac{2 k_{\parallel}}{S}}\, \mathrm{e}^{\mathrm{i} \mathbf{k}_{\parallel} \cdot \mathbf{r}_{\parallel}} \, \mathrm{e}^{-k_{\parallel} | z |},
\end{equation} 
where $S$ denotes the in-plane area, and the separation index $\mu$ corresponds to the in-plane wavevector $\mathbf{k}_{\parallel}$. The function given in Eq.~\eqref{eq : Phi Plan} exhibits an evanescent profile in the $z$ direction, with a penetration depth $1/k_{\parallel}$ that reflects the breaking of translational symmetry along $z$. Substituting Eq.~\eqref{eq : Phi Plan} into Eq.~\eqref{eq : lambda phi} with $\mathbf{N} \equiv \mathbf{\hat{z}}$, one finds $\lambda = \varepsilon_2 + \varepsilon_{\infty} - 1$ for all $\mathbf{k}_{\parallel}$. From Eq.~\eqref{surf_plasm_freq}, the surface plasmon frequency follows as
\begin{equation}
    \label{eq : omega sp}
    \omega_{\mathrm{sp}} = \frac{\omega_{\mathrm{p}}}{\sqrt{\varepsilon_{\infty} + \varepsilon_2}},
\end{equation}
which is independent of $\mathbf{k}_{\parallel}$ because $\lambda$ is the same for all $\mathbf{k}_{\parallel}$. For simplicity, we consider an air-metal interface with $\varepsilon_2 = 1$ and $\varepsilon_{\infty} = 1$, yielding the well-known result $\omega_{\mathrm{sp}} = \omega_{\mathrm{p}}/\sqrt{2}$. Since the planar interface can be regarded as the $a \to \infty$ limit of a metallic sphere, it is noteworthy that the planar surface plasmon frequency can be recovered from the nanosphere result by taking $l \to + \infty$ in Eq.~\eqref{eq : omega l}. Thus, surface plasmons of the planar interface can be interpreted as the asymptotic limit of high-order multipolar modes of a metallic sphere.  

Unlike the case of LSP polaritons in nanoparticles, where the quasistatic regime dominates because the polariton wavelength $\sim c/\omega_{\mathrm{p}}$ is much larger than the nanoparticle radius $a$, here we aim to capture the full PSP polariton dispersion using our quantum description. Specifically, we are interested in the dependence of the PSP polariton energy on the in-plane wavevector modulus $k_{\parallel}$. In the quasistatic regime, where the polariton wavelength $\sim c/\omega_{\mathrm{p}}$ is much larger than the penetration depth $1/k_{\parallel}$, propagation effects are negligible; however, they become essential for $c/\omega_{\mathrm{p}} \lesssim 1/k_{\parallel}$. In the fully retarded regime $c/\omega_{\mathrm{p}} \ll 1/k_{\parallel}$, PSP polaritons behave as purely photonic excitations extending throughout the dielectric half-space~\cite{Greffet}.  

To incorporate propagation effects, we introduce a generalized penetration depth $1/\gamma(k_{\parallel})$, which ensures the correct asymptotic behavior: It reduces to $1/k_{\parallel}$ in the limit $k_{\parallel} \gg \omega_{\mathrm{p}}/c$, while diverging inside the dielectric medium for $k_{\parallel} \ll \omega_{\mathrm{p}}/c$. Consequently, we replace $k_{\parallel} \to \gamma (k_{\parallel})$ in both the normalized factor and the $z$-dependent part of the quasistatic mode functions defined in Eq.~\eqref{eq : Phi Plan}, yielding
\begin{equation}
\label{eq : Phi Plan2}
g_{\mathbf{k}_{\parallel}}(\mathbf{r}) = \sqrt{\frac{2 \gamma (k_{\parallel})}{S}}\, \mathrm{e}^{\mathrm{i} \mathbf{k}_{\parallel} \cdot \mathbf{r}_{\parallel}} \,\mathrm{e}^{-\gamma (k_{\parallel}) | z |}.
\end{equation}
We emphasize that Eq.~\eqref{eq : Phi Plan2} accounts only for evanescent decay in the regime of the normal skin effect. When nonlocal effects are included, the electric field may decay more slowly, as shown in Ref.~\cite{Larkin2017}.
Introducing the decomposition $\gamma(k_{\parallel}) = \Theta(-z) \gamma_{1}(k_{\parallel}) + \gamma_{2}(k_{\parallel}) \Theta(z)$, we obtain using Eq.~\eqref{effective_length} the quantization length 
\begin{equation}
L_{k_{\parallel}} = \frac{1}{\sqrt{k^2_{\parallel}+ \gamma^{2}_{1} (k_{\parallel})}}.
\end{equation}
Accordingly, the polarization field including propagation effects, derived from Eq.~\eqref{P_elec_bos}, reads
\begin{align}
\mathbf{P}(\mathbf{r})
=& \sum_{\mathbf{k}_{\parallel}} 
\sqrt{\frac{\hbar \omega_{\mathrm{p}}\varepsilon_{0}\gamma_{1}(k_{\parallel}) L^{2}_{k_{\parallel}}}
{ S}} \Theta(-z) 
\left( B^{\phantom{\dagger}}_{\mathbf{k}_{\parallel}} + B^{\dagger}_{-\mathbf{k}_{\parallel}} \right)
\nonumber \\
& \times \left[ \mathrm{i} \mathbf{k}_{\parallel} + \gamma_{1}(k_{\parallel}) \mathbf{\hat{z}} \right]
\mathrm{e}^{\mathrm{i} \mathbf{k}_{\parallel} \cdot \mathbf{r}_{\parallel}} \,
\mathrm{e}^{\gamma_{1}(k_{\parallel}) z},
\label{pola_quant_plan}
\end{align}
which couples only to TM photon modes [Eq.~\eqref{eq:TE_TM}]. Together with Eqs.~\eqref{H_elec_bos} and \eqref{eq : Hph quantum}, and writing the quantization volume as $V_{0}=L_z S$, the total Hamiltonian is
\begin{align}
\label{full_H_plane}
H =&\; \hbar \omega_{\mathrm{p}} \sum_{\mathbf{k}_{\parallel}} 
B_{\mathbf{k}_{\parallel}}^{\dagger} B^{\phantom{\dagger}}_{\mathbf{k}_{\parallel}}
+ \sum_{\mathbf{k}} \hbar c k \, a_{\mathbf{k}}^{\dagger} a^{\phantom{\dagger}}_{\mathbf{k}} \nonumber \\
& + \mathrm{i} \hbar \sum_{\mathbf{k}} C_{\mathbf{k}}
\left( B^{\phantom{\dagger}}_{-\mathbf{k}_{\parallel}} + B_{\mathbf{k}_{\parallel}}^{\dagger} \right)
\left( a^{\phantom{\dagger}}_{\mathbf{k}} - a^{\dagger}_{-\mathbf{k}} \right),
\end{align}
with the light-matter coupling strength
\begin{equation}
C_{\mathbf{k}} = \sqrt{\frac{\omega_{\mathrm{p}}ck \gamma_{1}(k_{\parallel}) L^{2}_{k_{\parallel}}}{2 L_z}} \left(\frac{k_{\parallel}}{k} \right),
\label{eq:C_k_PSP}
\end{equation}
and $\mathbf{k}=(\mathbf{k}_{\parallel},k_{z})$. Owing to in-plane translational invariance, the Hamiltonian \eqref{full_H_plane} factorizes as $H=\sum_{\mathbf{k}_{\parallel}} H_{\mathbf{k}_{\parallel}}$, with $[H_{\mathbf{k}_{\parallel}},H_{\mathbf{k}'_{\parallel}}]=0$. For a fixed in-plane wavevector $\mathbf{k}_{\parallel}$, the Hamiltonian~\eqref{full_H_plane} retains the same structure as Eq.~\eqref{full_H_sphere} describing LSP polaritons. In contrast to the latter, however, PSP polariton modes with frequencies below the light cone are decoupled from the radiative continuum and therefore do not experience radiative damping; they can thus be regarded as genuine eigenmodes of the system. Each block can then be diagonalized as $H_{\mathbf{k}_{\parallel}} = \hbar \Omega_{k_{\parallel}} p_{\mathbf{k}_{\parallel}}^{\dagger} p_{\mathbf{k}_{\parallel}}^{\phantom{\dagger}}$ using a Hopfield-Bogoliubov transformation. The PSP polariton annihilation operator is
\begin{equation}
\label{eq : pol plan}
p_{\mathbf{k}_{\parallel}} = w_{\mathbf{k}_{\parallel}} B^{\phantom{\dagger}}_{\mathbf{k}_{\parallel}} + x_{\mathbf{k}_{\parallel}} B_{\mathbf{k}_{\parallel}}^{\dagger}+ \sum_{k_{z}} \left( y_{\mathbf{k}} a^{\phantom{\dagger}}_{\mathbf{k}} + z_{\mathbf{k}} a^{\dagger}_{\mathbf{k}} \right),
\end{equation}
with Hopfield coefficients satisfying the normalization
\begin{equation}
\label{eq : norm_Hopfield}
\vert w_{\mathbf{k}_{\parallel}} \vert^2 - \vert x_{\mathbf{k}_{\parallel}} \vert^2 + \sum_{k_{z}} \left( \vert y_{\mathbf{k}} \vert^2 - \vert z_{\mathbf{k}} \vert^2 \right) = 1.
\end{equation}

The Heisenberg equation of motion $[p_{\mathbf{k}_{\parallel}}, H_{\mathbf{k}_{\parallel}}] = \hbar \Omega_{k_{\parallel}} p_{\mathbf{k}_{\parallel}}$ yields the coupled system
\begin{subequations}
\label{eq : Hopfield 1_plane}
\begin{align}
w_{\mathbf{k}_{\parallel}} \left( \omega_{\mathrm{p}} - \Omega_{k_{\parallel}} \right)
&= +\mathrm{i} \sum_{k_{z}} C_{\mathbf{k}} \left( y_{\mathbf{k}} + z_{\mathbf{k}} \right), \\
x_{\mathbf{k}_{\parallel}} \left( \omega_{\mathrm{p}} + \Omega_{k_{\parallel}} \right)
&= -\mathrm{i} \sum_{k_{z}} C_{\mathbf{k}} \left( y_{\mathbf{k}} + z_{\mathbf{k}} \right), \\
y_{\mathbf{k}} \left( c k - \Omega_{k_{\parallel}} \right)
&= \mathrm{i} C_{\mathbf{k}} \left( x_{\mathbf{k}_{\parallel}} - w_{\mathbf{k}_{\parallel}} \right), \\
z_{\mathbf{k}} \left( c k + \Omega_{k_{\parallel}} \right)
&= \mathrm{i} C_{\mathbf{k}} \left( x_{\mathbf{k}_{\parallel}} - w_{\mathbf{k}_{\parallel}} \right),
\end{align}
\end{subequations}
which after eliminating the Hopfield coefficients provides the eigenvalue equation
\begin{equation}
\label{eq : eigenvalue spp quantum}
    \Omega_{k_{\parallel}}^2 = \omega_{\mathrm{p}}^2 - \sum_{k_z} \frac{4 c k \omega_{\mathrm{p}} C_{\mathbf{k}}^2}{c^2 k^2 - \Omega_{k_{\parallel}}^2}.
\end{equation}

The solutions of Eq.~\eqref{eq : eigenvalue spp quantum} yield the PSP polariton frequencies $\Omega_{k_{\parallel}}$, which depend on the effective penetration depth $1/\gamma(k_{\parallel})$ introduced above through the light-matter coupling strength $C_{\mathbf{k}}$. Our goal is now to derive an equation that allows one to determine the penetration depth self-consistently with the PSP polariton frequencies. Decomposing the electric field as $\mathbf{E}(\mathbf{r},t)=\sum_{\mathbf{k}_{\parallel}} \mathbfcal{E}_{\mathbf{k}_{\parallel}} (\mathbf{r},t)$, where $\mathbfcal{E}_{\mathbf{k}_{\parallel}} (\mathbf{r},t)$ evolves in time as $ \mathrm{e}^{-\mathrm{i} \Omega_{k_{\parallel}} t}$, the wave equation takes the form~\cite{Maier}
\begin{equation}
\label{eq : wave equation}
\nabla^2 \mathbfcal{E}_{\mathbf{k}_{\parallel}} (\mathbf{r},t)= \frac{\varepsilon(\Omega_{k_{\parallel}})}{c^2} \frac{\partial^2 \mathbfcal{E}_{\mathbf{k}_{\parallel}} (\mathbf{r},t)}{\partial t^2},
\end{equation}
where $\varepsilon(\Omega_{k_{\parallel}}) = \Theta(-z) \varepsilon_1(\Omega_{k_{\parallel}}) + \Theta(z)\varepsilon_2$, and where we have used that the electric field is divergence-free away from the interface, i.e., $\bm{\nabla} \cdot \mathbf{E}=0$ for $\mathbf{r} \notin \Sigma$. Moreover, the Maxwell-Faraday equation implies that the electric field is irrotational. Using $\mathbf{E}_{\parallel} = - \nabla \phi$ together with the results of Appendix \ref{app_transv_long}, one finds that the total electric field satisfies $\mathbf{E} \propto \nabla \phi$, where $\phi$ is expanded over the functions introduced in Eq.~\eqref{eq : Phi Plan2}. The wave equation \eqref{eq : wave equation} then yields the dispersion relation in each medium
\begin{subequations}
\label{eq : Helmholtz 1}
\begin{align}
\varepsilon_1(\Omega_{k_{\parallel}}) \frac{\Omega_{k_{\parallel}}^2}{c^2}
&= k_{\parallel}^2 - \gamma^{2}_1(k_{\parallel}), \qquad z \leqslant 0, \\
\frac{\Omega_{k_{\parallel}}^2}{c^2}
&= k_{\parallel}^2 - \gamma^{2}_2(k_{\parallel}), \qquad z > 0.
\end{align}
\end{subequations}
In Appendix \ref{app_planar_interf}, we show that the only solution of Eq.~\eqref{eq : eigenvalue spp quantum} that is compatible with Eqs.~\eqref{eq : Helmholtz 1} corresponds to a real positive frequency lying below the light cone and below the plasma frequency, i.e., satisfying $\Omega_{k_{\parallel}} < c k_{\parallel}$,  $\Omega_{k_{\parallel}} < \omega_{\mathrm{p}}$, namely
\begin{equation}
    \label{eq : Disp SPP}
    \Omega_{k_{\parallel}} = \sqrt{c^2 k_{\parallel}^2 + \omega^2_{\mathrm{sp}} - \sqrt{c^4 k_{\parallel}^4 + \omega^4_{\mathrm{sp}}}},
\end{equation}
which coincides with the PSP polariton dispersion obtained from Maxwell's equations~\cite{Maier, Greffet, Pitarke}. This result shows the intrinsically nonperturbative nature of the light-matter coupling in planar metal-dielectric interfaces.

\begin{figure}[t]
\begin{center}
\includegraphics[scale=0.3]{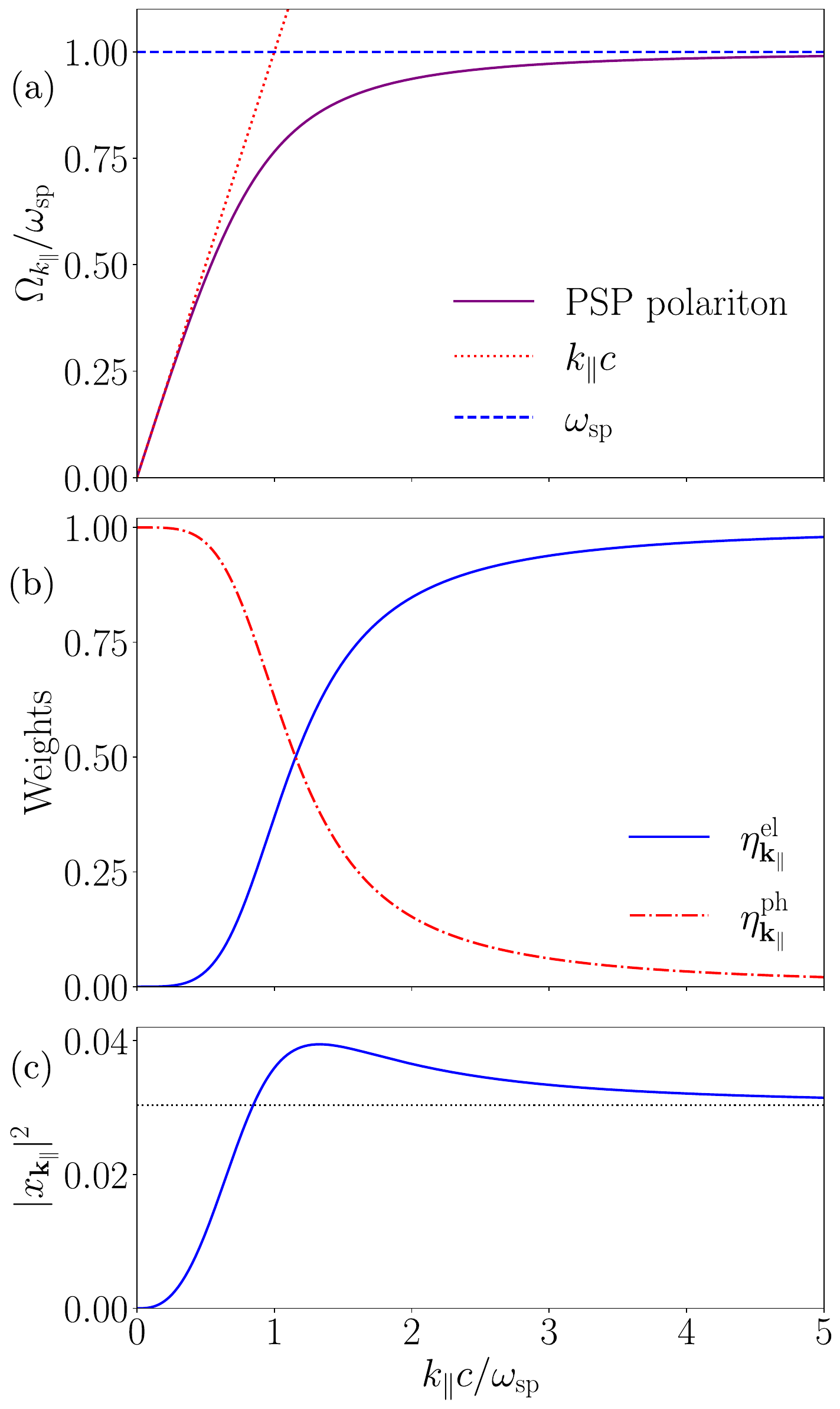}
\caption {(a) PSP polariton dispersion $\Omega_{k_{\parallel}}$ from Eq.~\eqref{eq : Disp SPP} (purple). The light line $ck_{\parallel} $ is in red (dotted), and the surface plasmon frequency $\omega_\mathrm{sp}$ is shown as a blue dashed line. (b) Electronic (blue, solid) and photonic (red, dash-dotted) weights of PSP polaritons, $\eta_{\mathbf{k}_{\parallel}}^{\mathrm{el}}$ and $\eta_{\mathbf{k}_{\parallel}}^{\mathrm{ph}}$. (c) Square modulus of the anomalous Hopfield coefficient, $\vert x_{\mathbf{k}_{\parallel}} \vert^2$. The dotted line represents the asymptotic value in the quasistatic limit, $k_{\parallel} \gg \omega_{\mathrm{p}}/c$, which corresponds to the plasmon population in the ground state.} 
\label{fig : weights spp}
\end{center}
\end{figure}

The PSP polariton dispersion is shown in Fig.~\ref{fig : weights spp}(a) and using Eqs.~\eqref{eq : Helmholtz 1} and \eqref{eq : Disp SPP}, the inverse of the penetration depths in the metal and in the dielectric are given by 
\begin{equation}
    \label{eq : gamma 1}
    \gamma_{1}(k_{\parallel}) = \frac{\omega_{\mathrm{sp}}}{c} \sqrt{1 + \sqrt{1 +{c^4k_{\parallel}^4 / \omega_{\mathrm{sp}}^4}}}
\end{equation}
and
\begin{equation}
    \label{eq : gamma 2}
    \gamma_{2}(k_{\parallel}) = \frac{\omega_{\mathrm{sp}}}{c} \sqrt{-1 + \sqrt{1 +{c^4k_{\parallel}^4 / \omega_{\mathrm{sp}}^4}}}.
\end{equation}
In the quasistatic limit $k_{\parallel} \gg \omega_{\mathrm{p}}/c$, one finds $\Omega_{k_{\parallel}} \to \omega_{\mathrm{sp}}$, with $\gamma_1(k_{\parallel}) \sim \gamma_2(k_{\parallel}) \sim k_{\parallel}$, recovering the Laplace penetration depth. Conversely, in the fully retarded regime $k_{\parallel} \ll \omega_{\mathrm{p}}/c$, one finds $\Omega_{k_{\parallel}} \sim k_{\parallel} c$, with $\gamma_1(k_{\parallel}) \sim \omega_{\mathrm{p}}/c$, corresponding to the skin depth, while $\gamma_2(k_{\parallel}) \to 0$, consistent with free photons propagating throughout the dielectric half-space. 

Moreover, the photonic and electronic content of PSP polaritons can be deduced from the Hopfield coefficients. As in Sec.~\ref{sec : LSP qs}, we introduce the electronic weight 
\begin{equation}
\label{eq:electronic_weight_def}
\eta_{\mathbf{k}_{\parallel}}^{\mathrm{el}}= |w_{\mathbf{k}_{\parallel}} |^2 - |x_{\mathbf{k}_{\parallel}}|^2
\end{equation}
and the photonic weight $\eta_{\mathbf{k}_{\parallel}}^{\mathrm{ph}} = \sum_{k_z} ( | y_{\mathbf{k}} |^2 - |z_{\mathbf{k}} |^2 )$, which obey the normalization condition $\eta_{\mathbf{k}_{\parallel}}^{\mathrm{el}} + \eta_{\mathbf{k}_{\parallel}}^{\mathrm{ph}} = 1$ stemming from Eq.~\eqref{eq : norm_Hopfield}. A calculation detailed in Appendix \ref{app_planar_interf} shows that the electronic weight can be written as
\begin{equation}
    \label{eq : M k result}
     \eta_{\mathbf{k}_{\parallel}}^{\mathrm{el}} = \frac{\left[k_{\parallel}^2 + \gamma^{2}_{1} (k_{\parallel}) \right] \gamma^{3}_{2} (k_{\parallel})}{\left[k_{\parallel}^2 + \gamma^{2}_{1}  (k_{\parallel}) \right] \gamma^{3}_{2} (k_{\parallel}) + \gamma_{1} (k_{\parallel})\omega^2_{\mathrm{sp}} k_{\parallel}^2/c^2}.
\end{equation}
The electronic and photonic weights are plotted in Fig.~\ref{fig : weights spp}(b). As expected, PSP polaritons are (asymptotically) fully photonic in the retarded regime $k_{\parallel} \ll \omega_{\mathrm{p}}/c$, purely plasmonic in the quasistatic regime $k_{\parallel} \gg \omega_{\mathrm{p}}/c$, and a plasmon-photon mixture in between.

The combination of Eqs.~\eqref{eq : norm_Hopfield} and \eqref{eq : Hopfield 1_plane} yields relations among the Hopfield coefficients, from which we obtain
\begin{equation}
    \label{eq : anomalous Hopfield}
    \vert x_{\mathbf{k}_{\parallel}}\vert^2 
    =  \frac{\left( \omega_{\mathrm{p}}- \Omega_{k_{\parallel}}\right)^2}{4 \omega_{\mathrm{p}} \Omega_{k_{\parallel}}} 
    \, \eta_{\mathbf{k}_{\parallel}}^{\mathrm{el}},
\end{equation}
as represented in Fig.~\ref{fig : weights spp}(c) in the case of $\varepsilon_\infty=\varepsilon_2=1$, using the expressions for the dispersion relation and the electronic weight provided in Eqs.~\eqref{eq : Disp SPP} and~\eqref{eq : M k result}. The anomalous contribution to the electronic weight given in Eq.~\eqref{eq : anomalous Hopfield} is non-negligible, highlighting that the coupled plasmon-photon system in the PZW representation, which gives rise to PSP polaritons, operates intrinsically in the ultrastrong coupling regime. As in Sec.~\ref{sec : LSP qs}, it is instructive to evaluate the bulk plasmon population in the polaritonic ground state, $\langle G_{\mathbf{k}_{\parallel}} \vert B_{\mathbf{k}_{\parallel}}^{\dagger} B_{\mathbf{k}_{\parallel}}^{\phantom{\dagger}} \vert G_{\mathbf{k}_{\parallel}} \rangle$, in the quasistatic limit $k_{\parallel} \gg \omega_{\mathrm{p}}/c$. In this regime, the electronic weight approaches unity [see Fig.~\ref{fig : weights spp}(b)], and using Eq.~\eqref{eq : pol plan} it follows that $B_{\mathbf{k}_{\parallel}} = w_{\mathbf{k}_{\parallel}}^{*} p_{\mathbf{k}_{\parallel}}^{\phantom{\dagger}} - x_{\mathbf{k}_{\parallel}}^{\phantom{*}} p_{\mathbf{k}_{\parallel}}^{\dagger}$. Consequently, $\langle G_{\mathbf{k}_{\parallel}} \vert B_{\mathbf{k}_{\parallel}}^{\dagger} B_{\mathbf{k}_{\parallel}}^{\phantom{\dagger}} \vert G_{\mathbf{k}_{\parallel}} \rangle = \vert x_{\mathbf{k}_{\parallel}} \vert^2$. 
Using Eq.~\eqref{eq : anomalous Hopfield} together with $\Omega_{k_{\parallel}} = \omega_{\mathrm{sp}}$ in the quasistatic limit, we obtain
    $\langle G_{\mathbf{k}_{\parallel}} \vert B_{\mathbf{k}_{\parallel}}^{\dagger} B_{\mathbf{k}_{\parallel}}^{\phantom{\dagger}} \vert G_{\mathbf{k}_{\parallel}} \rangle 
    = 3/4 \sqrt{2} - 1/2
    \simeq 3.0\times10^{-2}$,
which corresponds to the asymptotic limit $l\gg1$ for a particle or a cavity as discussed in Sec.~\ref{sec : LSP qs} [see Fig.~\ref{fig : GSPop}(a)]. As discussed in the case of the nanoparticle, this nonzero but relatively small plasmon population in the ground state can be significantly enhanced by increasing the dielectric contrast $\varepsilon_2$ and the high-frequency screening constant $\varepsilon_\infty$ of the system [see Fig.~\ref{fig : GSPop}(b)].

\section{Conclusion}
\label{conclusion}

We have developed a microscopic quantum description of surface plasmon polaritons at arbitrary metal-dielectric interfaces within the PZW representation. In particular, we introduced a quantization procedure for the polarization field based on the resolution of the Laplace equation. By coupling the polarization to the displacement field of the free EM field, we obtained a general Hamiltonian formulation of surface plasmon polaritons.

We applied this framework to the case of a spherical metallic nanoparticle and successfully recovered the LSP modes. Analytical expressions valid in the quasistatic regime for both the frequency shift and the decay rate were obtained and shown to be in excellent agreement with classical predictions. We also applied our formalism to propagating surface plasmon polaritons at a planar metal-dielectric interface. By considering the full Hamiltonian, we recovered the classical dispersion relation of PSPs and provided their photonic and electronic contents. The method of quantization shown here can be generalized to the case of layered geometries \cite{Economou_1969}, which are of interest for nanophotonic applications. In the quasistatic limit, we have shown that the bulk plasmon population in the polaritonic ground state is finite, implying that plasmon-photon systems at metal-dielectric interfaces intrinsically operate in the ultrastrong coupling regime. As a corollary, these systems display anomalous quantum fluctuations governed exclusively by the interface geometry and the refractive indices of the media.

An interesting question concerns the precise relation between the PZW approach and the minimal-coupling representation. Both configurations considered in this work have been analyzed previously within the minimal-coupling framework~\cite{Allard,Alpeggiani}. In the microscopic treatment of Ref.~\cite{Allard}, addressing metallic nanoparticles, the $\mathbf{A}^2$ term was neglected, whereas in Ref.~\cite{Alpeggiani} this term was explicitly retained, yielding (though through a cumbersome derivation) the correct PSP dispersion relation [Eq.~\eqref{eq : Disp SPP}]. By construction, the PZW Hamiltonian $H$ can be obtained from the minimal-coupling Hamiltonian $H_{\mathrm{mc}}$ via the unitary transformation $H = T^{\dagger} H_{\mathrm{mc}}T$, with $T = \exp\left(-\tfrac{\mathrm{i}}{\hbar}\int \mathrm{d}^3\mathbf{r} \, \mathbf{A}(\mathbf{r})\cdot \mathbf{P}(\mathbf{r})\right)$. However, the unitarity of this transformation requires a complete basis of the Hilbert space, a condition not satisfied in our analysis. For the nanoparticle, the polarization field was truncated to its dipolar component, while for the planar interface only exponentially decaying solutions of Laplace's equation were retained, with unphysical growing solutions discarded. Consequently, the basis sets in both cases are incomplete, so the two Hamiltonians are not related by a unitary transformation. Instead, one obtains independent derivations that nevertheless converge to the correct dispersion relation. We here demonstrate that the PZW formalism provides a more general, direct, and conceptually transparent route to this outcome.

A natural extension of this work is the inclusion of nonradiative losses, which can be incorporated either within the Green's function formalism or through Fano diagonalization~\cite{Fano1961}. For instance, Ohmic losses in the metal may be described by adopting a complex dielectric function, $\varepsilon_{1}(\omega) = \varepsilon_\infty - \omega_{\mathrm{p}}^2/(\omega^2 + \mathrm{i}\tau^{-1}\omega)$, where $\tau^{-1}$ denotes the damping rate. To maintain consistency between Maxwell's equations and the microscopic Hamiltonian formulation, the system must then be coupled to an external bath of harmonic oscillators~\cite{Huttner1992} with an Ohmic spectral density $J(\omega) \propto \tau^{-1} \omega$~\cite{Leggett,Ingold1988}. In this case, our approach would effectively converge toward macroscopic QED.

The theory presented here enables further investigations of nanophotonic systems within a rigorous quantum framework. A first case of interest is the interaction between a single two-level system and a metallic nanoparticle, a configuration that has been studied at the classical level to investigate the enhancement of spontaneous emission arising from the nanoparticle's antenna effect~\cite{Kuhn2006}. Our model would provides a fully quantum-mechanical description of the optical properties of the emitted radiation in such experiments. Moreover, our description basically reduces to a Rabi Hamiltonian emulating a qubit-circuit or atom-cavity systems that are well established at other frequency scales~\cite{Haroche_book}. This analogy opens promising perspectives for exploring nonclassical states in nanoplasmonic systems operating in the visible range, provided appropriate experimental conditions such as cryogenic temperatures can be achieved. Another relevant scenario concerns coupled metallic nanoparticles or parallel metallic plates, for which the PZW representation may offer a particularly transparent framework to describe Casimir-Polder interactions~\cite{Barrera_2006}. Such coupled systems also support superradiant and subradiant collective states, and our formulation could prove valuable in elucidating their quantum-optical properties.

\begin{acknowledgments}

We thank Didier Felbacq for insightful discussions. This work of the Interdisciplinary Thematic Institute QMat, as part of the ITI 2021-2028 program of the University of Strasbourg, CNRS and Inserm, was supported by IdEx Unistra (ANR 10 IDEX 0002), and by SFRI STRAT'US project (ANR 20 SFRI 0012) and EUR QMAT ANR-17-EURE-0024 under the framework of the French Investments for the Future Program. This work benefited also from State support managed by the National Research Agency under the France 2030 program, referenced by ANR‐22‐CMAS‐0001, as well as ERC-COG-863487 “UNIQUE”.

\end{acknowledgments}

\appendix


    
\section{Proportionality of the longitudinal and transverse polarizations and sum rule for plasmon frequencies}
\label{app_transv_long}

In this Appendix, we demonstrate that the property of the polarization field being both curl-free and divergence-free implies the proportionality between the transverse and longitudinal components of the polarization. 

The polarization field $\mathbf{P}$ can be expressed either as the gradient of a scalar function $h$ or as the curl of a vector field $\mathbf{V}$, i.e., $\mathbf{P} = \nabla h$ or $\mathbf{P} = \nabla \times \mathbf{V}$. The boundary conditions imposed by the considered geometry uniquely determine both \(h\) and \(\mathbf{V}\), and hence \(\mathbf{P}\) is also unique. Therefore, one has $\nabla h = \nabla \times \mathbf{V}$. The polarization can thus be written as a linear combination of its irrotational and solenoidal components:
\begin{equation}
\mathbf{P} = \beta \nabla h + (1 - \beta) \nabla \times \mathbf{V}, \quad \beta \in \mathbb{R}.
\end{equation}
This corresponds to a Helmholtz decomposition, $\mathbf{P} = \mathbf{P}_{\parallel} + \mathbf{P}_{\perp}$, with \(\mathbf{P}_{\parallel} = \beta \nabla h\) and \(\mathbf{P}_{\perp} = (1 - \beta) \nabla \times \mathbf{V}\). From the uniqueness of \(\mathbf{P}\), it follows that \(\mathbf{P}_{\perp}\) and \(\mathbf{P}_{\parallel}\) must be proportional,
\begin{equation}
\label{pola_appendix}
\mathbf{P}_{\perp} = \frac{1 - \beta}{\beta} \, \mathbf{P}_{\parallel} \equiv \lambda \, \mathbf{P}_{\parallel},
\end{equation}
where \(\lambda = (1 - \beta)/\beta\). This yields a proportionality relation between the transverse and longitudinal parts of the polarization field. Since the general solution to Laplace's equation \eqref{eq : Phi Laplace} can be expressed as a superposition of independent harmonic functions, each associated with a mode \(\mu\), the proportionality relation \eqref{pola_appendix} thus remains valid for every \(\mu\). Expanding the transverse polarization in analogy with the longitudinal polarization \eqref{eq : general P Laplace},
$\mathbf{P}_{\perp} = \sum_{\mu} \mathbf{P}_{\perp,\mu}$, the total polarization reads $\mathbf{P} = \sum_{\mu} \mathbf{P}_{\mu} = \sum_{\mu} ( \mathbf{P}_{\parallel,\mu} + \mathbf{P}_{\perp,\mu})$. Thus, for each \(\mu\), Eq.~\eqref{pola_appendix} generalizes to
\begin{equation}
\label{eq : P lambda2}
\mathbf{P}_{\perp,\mu} =
\begin{cases}
 - \mathbf{P}_{\parallel,\mu}, & \mathbf{r} \in \mathbb{R}^3 \setminus V, \\
\lambda_{\mu} \, \mathbf{P}_{\parallel,\mu}, & \mathbf{r} \in V,
\end{cases}
\end{equation}
where we have used the fact that the total polarization vanishes inside the dielectric.

\begin{figure}[t]
\includegraphics[width = \columnwidth]{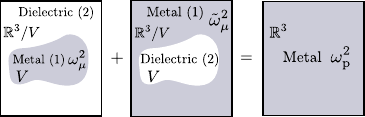}
\caption {Schematic illustration of the plasmonic frequency sum rule: Representation of the metallo-dielectric interface, as shown in Fig.~\ref{fig : Interface}, supporting a plasmonic mode with frequency $\omega_{\mu}$. Its complementary structure, obtained by interchanging the metallic and dielectric regions, is characterized by the plasmonic frequency $\tilde{\omega}_{\mu}$.} 
\label{fig : compl}
\end{figure}

We can use the above result to verify the sum rule for plasmon frequencies. To do so, we consider the complementary system, in which the metallic and dielectric regions are interchanged, as it is shown in Fig.~\ref{fig : compl}. The longitudinal polarization field is the same for both systems because the charge distribution remains the same. However, the total polarization differs. Hence, we can show that, similar to Eq.~\eqref{eq : P lambda2}, the transverse polarization $\tilde{\mathbf{P}}_{\perp,\mu}$ for the complementary system verifies
\begin{equation}
\label{eq : P lambda3}
\tilde{\mathbf{P}}_{\perp,\mu} =
\begin{cases}
 - \mathbf{P}_{\parallel,\mu}, & \mathbf{r} \in V, \\
 \tilde{\lambda}_{\mu} \, \mathbf{P}_{\parallel,\mu}, & \mathbf{r} \in \mathbb{R}^3 \setminus V,
\end{cases}
\end{equation}
where $\tilde{\lambda}_{\mu}$ is \textit{a priori} different from $\lambda_{\mu}$. Then, using the general result of Eq.~\eqref{eq : ortho P perp para}, which is still valid for the complementary system, yields
\begin{align}
\int_{V} \mathrm{d}^3 \mathbf{r} \,\mathbf{P}_{\parallel, \mu}^2 & =   \frac{1}{1 + \lambda_{\mu}} \int_{\mathbb{R}^3} \mathrm{d}^3 \mathbf{r} \,\mathbf{P}_{\parallel, \mu}^2 ,   \label{eq : P para 1} \\
 \int_{\mathbb{R}^3 \setminus V} \mathrm{d}^3 \mathbf{r} \,\mathbf{P}_{\parallel, \mu}^2 & = \frac{1}{1 + \tilde{\lambda}_{\mu}} \int_{\mathbb{R}^3} \mathrm{d}^3 \mathbf{r} \,\mathbf{P}_{\parallel, \mu}^2 .
\label{eq : P para 2}
\end{align}

Summing Eqs.~\eqref{eq : P para 1} and~\eqref{eq : P para 2} allows us to recover the integral of the longitudinal polarization over all space and yields 
\begin{equation}
    \label{eq : sum rule lamba}
    1 = \frac{1}{1 + \lambda_{\mu}} + \frac{1}{1 + \tilde{\lambda}_{\mu}},
\end{equation}
which, together with the definition of the frequencies given in Eq.~\eqref{surf_plasm_freq} gives the sum rule for surface plasmon frequencies
\begin{equation}
    \label{eq : sum rule freq}
    \omega_{\mu}^2 + \tilde{\omega}_{\mu}^2 = \omega_{\mathrm{p}}^2.
\end{equation}

\section{Eigenmodes of the metallic nanoparticle}
\label{app_met_nano}

In this Appendix, we determine the radiative corrections of the LSP resonance for a metallic nanoparticle in vacuum. This is achieved by computing the self-energy given in Eq.~\eqref{eq : Self energy}.

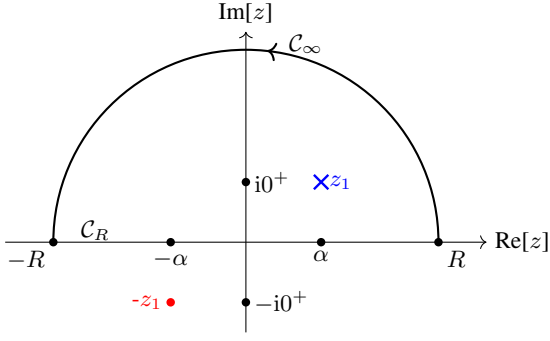
\begin{figure}[tb]
\centering
\begin{tikzpicture}[scale=0.8]
\draw[->] (-4,0) -- (4,0) node[right] {Re[$z$]};
\draw[->] (0,-1.5) -- (0,3.5) node[above] {Im[$z$]};
\draw[black, thick, postaction={decorate, decoration={markings, mark=at position -0.45 with {\arrow{<}}, }} ] (-3.2,0) arc (180:0:3.2);
\fill (3.2,0) circle (2pt) node[below right] {$R$};
\fill (-3.2,0) circle (2pt) node[below left] {$-R$};
\fill (0,1) circle (2pt) node[right] {$\mathrm{i} 0^{+}$};
\fill (0,-1) circle (2pt) node[right] {$-\mathrm{i} 0^{+}$};
\fill (1.25,0) circle (2pt) node[below] {$\alpha$};
\fill (-1.25,0) circle (2pt) node[below] {$-\alpha$};
\fill[red] (-1.25,-1) circle (2pt) node[left] {-$z_1$};

\draw[blue, line width=0.9pt] (1.25,1) ++(-0.12,-0.12) -- ++(0.24,0.24);
\draw[blue, line width=0.9pt] (1.25,1) ++(-0.12,0.12)  -- ++(0.24,-0.24);
\node at (1.25,1) [right, blue] {$z_1$};


\node at (1,3.3) {$\mathcal{C}_{\infty}$};
\node at (-2.5,0.2) {$\mathcal{C}_{R}$};
\end{tikzpicture}
\caption {Closed contour $\mathcal{C} = \mathcal{C}_{R} + \mathcal{C}_{\infty}$ that goes along the real axis from $-\infty$ to $+\infty$ ($\mathcal{C}_{R}$) and counterclockwise in the upper half complex plane along a semicircle centered at $0$ with a radius $R\rightarrow + \infty$ ($\mathcal{C}_{\infty}$). The integrand in Eq.~\eqref{int_1_eigenfreq} exhibits two poles at $\pm z_{1}$. The pole lying in the upper half-plane is displayed as a blue cross. } 
\label{fig : contour LSP}
\end{figure}

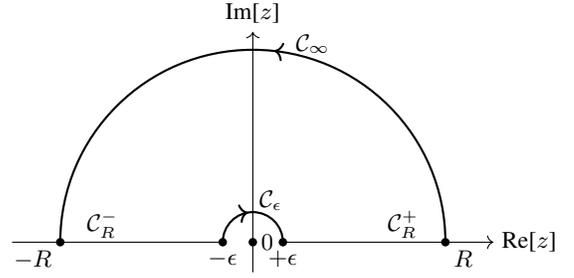
\begin{figure}[tb]
\centering
\begin{tikzpicture}[scale=0.8]
\draw[-] (-4,0) -- (-0.5,0) node[right]{ } ;
\draw[->] (0.5,0) -- (4,0) node[right] {Re[$z$]};
\draw[->] (0,-0.5) -- (0,3.5) node[above] {Im[$z$]};
\draw[black, thick, postaction={decorate, decoration={markings, mark=at position -0.45 with {\arrow{<}}, }} ] (-3.2,0) arc (180:0:3.2);
\draw[black, thick, postaction={decorate, decoration={markings, mark=at position 0.45 with {\arrow{>}}, }} ] (-0.5,0) arc (180:0:0.5);

\fill (3.2,0) circle (2pt) node[below right] {$R$};
\fill (-3.2,0) circle (2pt) node[below left] {$-R$};
\fill[black] (0,0) circle (2pt) node[right, black] {$0$};
\fill[black] (0.5,0) circle (2pt) node[below, black] {$+\epsilon$};
\fill[black] (-0.5,0) circle (2pt) node[below, black] {$-\epsilon$};
\node at (1,3.3) {$\mathcal{C}_{\infty}$};
\node at (-2.5,0.3) {$ \mathcal{C}_{{R}}^{-}$};
\node at (2.5,0.3) {$ \mathcal{C}_{{R}}^{+}$};
\node at (0.3,0.7) {$ \mathcal{C}_{\epsilon}$};
\end{tikzpicture}
\caption {Closed contour $\mathcal{C}' = \mathcal{C}_{{R}}^{-}+ \mathcal{C}_{\epsilon}+ \mathcal{C}_{{R}}^{+}+\mathcal{C}_{\infty}$, where $\mathcal{C}_{{R}}^{-}$ and $\mathcal{C}_{{R}}^{+}$ denote the real-axis paths going respectively from $-\infty$ to $-\epsilon \rightarrow 0^{-}$ and from $\epsilon \rightarrow 0^{+}$ to $+ \infty$, and $\mathcal{C}_{\epsilon}$ is a semicircle of radius $\epsilon$, centered at $0$ and going clockwise in the upper half complex plane.} 
\label{fig : contour LSP 2}
\end{figure}

Transforming the summation over $\mathbf{k}$ into an integral, and performing the angular integration, the self-energy reduces to
\begin{equation}
\label{eq : self energy integral}
\Sigma^{\mathrm{R}}(\Omega) = - \frac{\omega_{\mathrm{p}}}{\pi}\int_{-\infty}^{+\infty} \mathrm{d}x\; \frac{u^{2}(x) }{x^{2}-(\Omega  a/c + \mathrm{i} 0^{+})^{2}} , 
\end{equation}
where $u(x)\equiv \sin x/x - \cos x$. Let us consider the complex integral 
\begin{equation}
\label{int_1_eigenfreq}
\int_{\mathcal{C}}\mathrm{d}z\;  \frac{u^{2}(z) }{z^{2}-(\alpha + \mathrm{i} 0^{+})^{2}},  
\end{equation}
where $\alpha=\Omega a/c$ and $\mathcal{C}$ is the contour shown in Fig.~\ref{fig : contour LSP}. The integrand in Eq.~\eqref{int_1_eigenfreq} exhibits two poles at $z=\pm z_{1}=\pm \left(\alpha +\mathrm{i} 0^{+}\right)$. Noting that only the pole at $+z_{1}$ lies within the contour, the residue theorem yields
\begin{equation}
\int_{-\infty}^{+\infty} \mathrm{d}x\;\frac{u^{2}(x)}{x^{2}-z_{1}^{2}} + \int_{\mathcal{C}_{\infty}}\mathrm{d}z\; \frac{u^{2}(z)}{z^{2}-z_{1}^{2}} =\mathrm{i}\pi \frac{u^{2}(z_{1})}{z_{1}}.
\label{eq22}
\end{equation}
To evaluate the second term in Eq.~\eqref{eq22}, we first write 
\begin{equation}
\int_{\mathcal{C}_{\infty}} \mathrm{d}z\;\frac{u^{2}(z)}{z^{2}-z_{1}^{2}} \simeq \int_{\mathcal{C}_{\infty}} \mathrm{d}z\;\frac{u^{2}(z)}{z^{2}} + z^{2}_{1} \int_{\mathcal{C}_{\infty}} \mathrm{d}z\;\frac{u^{2}(z)}{z^{4}},
\label{eq2267}
\end{equation}
and then compute the two terms in the right-hand side of Eq.~\eqref{eq2267} using the Cauchy theorem on the contour $\mathcal{C}'$ shown in Fig.~\ref{fig : contour LSP 2}.

Exploiting the asymptotic behavior $u^{2}(z) \sim z^{4}/9$ for $z \to 0$, the two integrals over the contour $\mathcal{C}_{\epsilon}$ vanish, and therefore we obtain
\begin{align}
 \int_{\mathcal{C}_{\infty}}\mathrm{d}z\; \frac{u^{2}(z)}{z^{2}}
 &=   -\mathrm{p.v.} \int_{-\infty}^{+\infty} \mathrm{d}x\;\frac{u^{2}(x)}{x^{2}}
= -\frac{\pi}{3}, \label{eq32} \\
  \int_{\mathcal{C}_{\infty}} \mathrm{d}z\;\frac{u^{2}(z)}{z^{4}}
&=   -\mathrm{p.v.} \int_{-\infty}^{+\infty}\mathrm{d}x\; \frac{u^{2}(x)}{x^{4}}
= -\frac{2\pi}{15}.
\label{eq29}
\end{align}
Substituting Eqs.~\eqref{eq32} and \eqref{eq29} into Eq.~\eqref{eq22}, and using $u^{2}(z_{1}) \sim z^{4}_{1}/9$ as $\alpha \ll 1$, we obtain 
\begin{equation}
\int_{-\infty}^{+\infty} \mathrm{d}x\;\frac{u^{2}(x)}{x^{2}-z^{2}_{1}} \simeq \frac{\pi}{3} + \frac{2\pi}{15} z^{2}_{1}+\frac{\mathrm{i}\pi}{9} z^{3}_{1}.
\label{eq : result integral}
\end{equation}
Inserting Eq.~\eqref{eq : result integral} into Eq.~\eqref{eq : self energy integral} yields the self-energy
\begin{equation}
    \Sigma^{\mathrm{R}}(\Omega) = - \frac{\omega_{\mathrm{p}}}{3} - \frac{2\omega_{\mathrm{p}} \alpha^{2}}{15} - \mathrm{i}\frac{\omega_{\mathrm{p}} \alpha^{3}}{9}.
\end{equation}
Substituting this result into the Dyson equation~\eqref{eq : Dyson plasmon} and taking the imaginary part, the plasmon spectral function is obtained as
\begin{equation}
    \label{eq : spectral f expr}
    A(\Omega) = \frac{8 \omega_{\mathrm{p}}^{3} \alpha^{3} / 9}{\left( \Omega^{2} - \Omega_{0}^{2} \right)^{2} + \left( 2 \omega_{\mathrm{p}}^{2} \alpha^{3} / 9 \right)^{2}},
\end{equation}
where $\Omega_{0}^{2} = \omega_{\mathrm{M}}^{2} - 4 \omega_{\mathrm{p}}^{2} \alpha^{2} / 15$. The spectral function is peaked at $\Omega_{0}$, which can be approximated, to leading order in $\omega_{\mathrm{M}} a / c$, as
\begin{equation}
    \Omega_{0} \simeq \omega_{\mathrm{M}} \left( 1 - \frac{2\omega_{\mathrm{M}}^{2} a^{2}}{5c^{2}} \right).
\end{equation}
Introducing $\Omega_{0} = \omega_{\mathrm{M}} + \delta$, the corresponding radiative frequency shift is therefore given by Eq.~\eqref{eq : lamb shift}. Furthermore, in the vicinity of the resonance, the spectral function~\eqref{eq : spectral f expr} can be expressed in Lorentzian form as
\begin{equation}
    A(\Omega) \simeq \frac{\Gamma \sqrt{3}}{\left( \Omega - \Omega_{0} \right)^{2} + \left( \Gamma / 2 \right)^{2}},
\end{equation}
where the full width at half maximum is given, to lowest order in $\omega_{\mathrm{M}} a / c$, by Eq.~\eqref{eq : result gamma}.

\section{Eigenmodes of the planar interface}
\label{app_planar_interf}

\begin{figure}[tb]
\centering
\begin{tikzpicture}[scale=0.8]
\draw[->] (-4,0) -- (4,0) node[right] {Re[$k_z$]};
\draw[->] (0,-0.5) -- (0,3.5) node[above] {Im[$k_z$]};
\draw[black, thick, postaction={decorate, decoration={markings, mark=at position -0.45 with {\arrow{<}}, }} ] (-3.2,0) arc (180:0:3.2);
\fill (3.2,0) circle (2pt) node[below right] {$R$};
\fill (-3.2,0) circle (2pt) node[below left] {$-R$};
\fill (0,1) circle (2pt) node[right] {$\mathrm{i}\sqrt{k_{\parallel}^2 - \Omega_{k_{\parallel}}^2/c^2}$};

\node at (1,3.3) {$\mathcal{C}_{\infty}$};
\node at (-2.5,0.2) {$\mathcal{C}_R$};
\end{tikzpicture}
\caption {Closed contour $\mathcal{C} = \mathcal{C}_{R} + \mathcal{C}_{\infty}$ that goes along the real axis from $-\infty$ to $+\infty$ ($\mathcal{C}_{R}$) and counterclockwise in the upper half complex plane along a semicircle centered at $0$ with a radius $R\rightarrow + \infty$ ($\mathcal{C}_{\infty}$). The integrand in Eq.~\eqref{eq : eigenvalue passage} exhibits two poles at $\pm \mathrm{i}\sqrt{k_{\parallel}^2 - \Omega_{k_{\parallel}}^2/c^2}$. Only the pole lying in the upper half-plane is displayed. } 
\label{fig : contour SPP}
\end{figure}

Using Eq.~\eqref{eq:C_k_PSP} and going to the continuum limit, the eigenvalue equation \eqref{eq : eigenvalue spp quantum} can be put in the form
\begin{equation}
    \label{eq : eigenvalue passage}
        \Omega_{k_{\parallel}}^2 =  \omega_{\mathrm{p}}^2 \left[ 1 - \frac{ \gamma_1(k_{\parallel}) k_{\parallel}^2}{k_{\parallel}^2 + \gamma_1(k_{\parallel})^2} \frac{1}{\pi} \int_{-\infty}^{+\infty}  \frac{\mathrm{d}k_z}{k_z^2 + k_{\parallel}^2 - \Omega_{k_{\parallel}}^2/c^2} \right].
\end{equation}
Such an integral is calculated by going to the complex plane and using the contour $\mathcal{C}$ shown in Fig.~\ref{fig : contour SPP}. There is one pole inside the contour, located at $\mathrm{i}(k_{\parallel}^2 - \Omega_{k_{\parallel}}^2/c^2)^{1/2}$. Using the residue theorem, the eigenvalue equation \eqref{eq : eigenvalue passage} becomes
\begin{equation}
    \label{eq : Omega SPP Final}
        \Omega_{k_{\parallel}}^2 =  \omega_{\mathrm{p}}^2 \left[ 1 - \frac{ \gamma_1(k_{\parallel}) k_{\parallel}^2}{k_{\parallel}^2 + \gamma_1(k_{\parallel})^2} \frac{1}{\sqrt{k_{\parallel}^2 - \Omega_{k_{\parallel}}^2/c^2}} \right].
\end{equation}
Noticing that we are looking for solutions lying under the light-cone ($\Omega_{k_{\parallel}} < c k_{\parallel}$)
and that $\gamma_1(k_{\parallel}) \geqslant0$, Eq.~\eqref{eq : Omega SPP Final} implies that $\Omega_{k_{\parallel}}\leqslant \omega_{\mathrm{p}}$, imposing an upper limit for the eigenfrequency $\Omega_{k_{\parallel}}$. 
Using Eq.~\eqref{eq : Helmholtz 1} with $\varepsilon_{1}(\omega)$ given by Eq.~\eqref{eq : Drude} and $\varepsilon_{\infty} = 1$, the eigenvalue equation \eqref{eq : Omega SPP Final} then takes the form
\begin{equation}
    \left[  \Omega_{k_{\parallel}}^4 - 2 \Omega_{k_{\parallel}}^2 \left( \omega_{\mathrm{sp}}^2 + c^2 k_{\parallel}^2\right) + 2 \omega_{\mathrm{sp}}^2 c^2 k_{\parallel}^2\right] \mathcal{F}(\Omega_{k_{\parallel}}) = 0,
\end{equation}
where $\mathcal{F}(\Omega_{k_{\parallel}})$ is a sixth-order polynomial in $\Omega_{k_{\parallel}}$. The zeros of the function $\mathcal{F}(\Omega_{k_{\parallel}})$ yield solutions such that $\Omega_{k_{\parallel}} > c k_{\parallel}$ or $\Omega_{k_{\parallel}} > \omega_{\mathrm{p}}$, which is not compatible with the conditions specified above. Additionally, some solutions of the equation $\mathcal{F}(\Omega_{k_{\parallel}}) = 0$ are complex, and their real parts again lie above the light cone and above $\omega_{\mathrm{p}}$. Consequently, the zeros of the function $\mathcal{F}(\Omega_{k_{\parallel}})$ do not correspond to physical solutions of the system, and the dispersion relation of surface plasmon polaritons is obtained as the solution of the equation
\begin{equation}
      \Omega_{k_{\parallel}}^4 - 2 \Omega_{k_{\parallel}}^2 \left( \omega_{\mathrm{sp}}^2 + c^2 k_{\parallel}^2 \right) + 2 \omega_{\mathrm{sp}}^2 c^2 k_{\parallel}^2= 0.
\end{equation}
Among the four solutions of the above equation, only one of them corresponds to positive eigenfrequencies such that $\Omega_{k_{\parallel}} < c k_{\parallel}$. Therefore, the dispersion relation of surface plasmon polaritons is obtained as Eq.~\eqref{eq : Disp SPP}.

In addition to the dispersion relation discussed above, the electronic weight \eqref{eq:electronic_weight_def} can be readily obtained 
from Eqs.~\eqref{eq : norm_Hopfield} and \eqref{eq : Hopfield 1_plane} as $
    \eta_{\mathbf{k}_{\parallel}}^{\mathrm{el}} = 1/(1 + M_{\mathbf{k}_{\parallel}})$,
where
\begin{equation}
    \label{eq : M k}
    M_{\mathbf{k}_{\parallel}} = \frac{2 \omega_{\mathrm{sp}}^2 k_{\parallel}^2 \gamma_1(k_{\parallel})}{\pi c^2 [ k_{\parallel}^2 + \gamma_1^2(k_{\parallel}) ] } \int_{-\infty}^{+\infty}  \frac{\mathrm{d}k_z}{{( k_z^2 + k_{\parallel}^2 - \Omega_{k_{\parallel}}^2/c^2)}^2}.
\end{equation}
The integral in Eq.~\eqref{eq : M k} is computed similarly to the one in Eq.~\eqref{eq : eigenvalue passage}, using the contour of Fig.~\ref{fig : contour SPP}. We find
\begin{equation}
    \label{eq : M k result_}
    M_{\mathbf{k}_{\parallel}} = \frac{\omega_{\mathrm{sp}}^2 k_{\parallel}^2 \gamma_1(k_{\parallel})}{c^2 \left( k_{\parallel}^2 + \gamma_1^2(k_{\parallel}) \right) \left(k_{\parallel}^2 - \Omega_{k_{\parallel}}^2/c^2 \right)^{3/2}},
\end{equation}
which finally yields the electronic weight given in Eq.~\eqref{eq : M k result}.

\bibliography{Biblio}

\end{document}